%% file: NetSciX2017_Gidea_ARXIV.tex
\documentclass[11pt]{amsart}
\usepackage{epsfig}
\usepackage{epstopdf}
\usepackage{graphicx, subfigure}
\usepackage{amsmath, amssymb,mathabx,mathrsfs}
\usepackage{color}
\usepackage[toc,page]{appendix}
\usepackage{comment}

\theoremstyle{definition}

\theoremstyle{remark}

\numberwithin{equation}{section}

\theoremstyle{definition}

\newcommand{\eps}{\varepsilon}

\begin{document}


\title[Critical transitions in financial networks]{Topology data analysis of  critical transitions in financial networks}

\author{Marian\ Gidea$^\dag$}
\address{Yeshiva University, Department of Mathematical Sciences, New York, NY 10016, USA }
\email{Marian.Gidea@yu.edu}

\begin{abstract}{We develop a topology data analysis-based method  to detect early signs for critical transitions in financial data. From the time-series of multiple stock prices, we build time-dependent correlation networks, which  exhibit topological structures. We compute the persistent homology associated to these structures in order to track the changes in  topology  when approaching a  critical transition.
As a case study, we investigate a portfolio of  stocks during a period prior to the US financial crisis of 2007-2008, and show the presence of early signs of the critical transition.
}\end{abstract}

\maketitle

\section{Introduction}
\label{sec:introduction}

A \emph {critical transition} refers to an abrupt change  in the behavior
of a complex system --  arising due to  small changes in the external conditions --,
which makes the system switch from one steady state to some other steady state,
after undergoing a rapid transient process (e.g., `blue-sky catastrophe'
bifurcation). Examples of critical transitions are ubiquitous, including market crashes,  abrupt shifts in ocean circulation and climate,
regime changes in ecosystems, asthma
attacks and epileptic seizures, etc. A landmark paper on the theory of critical transitions and its applications
is \cite{Scheffer2009}.

A challenging problem of practical interest is to detect \emph{early signs} of critical transitions, that is, to identify
significant changes in the structure of the time-series data emitted by the  system prior to a sharp transition.
In this paper we propose  a new method to look for  critical transitions, based on measuring changes in the topological structure of the data. We consider systems that can be described as time-varying weighted networks, and we track  the changes in the topology of the network as the system approaches a critical transition.   We use  tools from \emph{topological data analysis}, more precisely  \emph{persistent homology}, to provide a precise characterization of the topology of the network throughout  its time-evolution.  We observe, in empirical data, that there are significant, measurable changes in the topology of the network as the underlying system approaches a critical transition.

The pipeline of our approach is the following. The \emph{input} of our procedure is a time-evolving weighted network $G(V,E)$, $w_t:E\to[0,\infty)$, i.e., a graph of nodes $V$ and edges $E$, with each edge $e\in E$ having assigned a weight $w_t(e)$ which varies in time. At each instant of time $t$, using a threshold value of the weight function as a parameter, we consider the threshold sub-network consisting of those edges whose weights are below  that threshold. We compute the homology of the clique complex determined by that sub-network. As we vary the threshold value,    some of the homology generators persist for a  large range of values while others disappear quickly. The persistent generators  provide   information about the significant, intrinsic patterns within the network, while the transient patterns may be redeemed as less significant or random. This information can be encoded  in terms of a  so-called  \emph{persistent diagram}, which provides a summary of the topological information on the network. As the time evolves, the topology of the network changes, and the corresponding persistent diagrams  also change. There is a natural metric (in fact, several) to measure distances between persistent diagrams.
It is important to note that \emph{persistent diagrams are robust}, meaning that small changes in the network yield persistent diagrams that are close to one another in terms of their mutual distances. The \emph{output} of our
procedure consists of a sequence of  distances measured between the persistent diagram at time $t$  and the persistent diagram at some initial time~$t_0$.

The salient features of our approach  are the following: \begin{enumerate}\item[(i)]     We process the input signal in its entirety, as we do not filter out noise from signal, \item[(ii)]  For   weighted networks,   we obtain a global  description of  all   threshold sub-networks, for all possible threshold values; \item[(iii)]
We describe in more  detail the structures of our  networks, unlike the statistical-type methods (e.g.,
centrality measures); \item[(iv)] We   provide an efficient way to compare weighted networks through the distances between the associated persistent diagrams, \item[(v)]  For time-dependent networks, we track the changes of the topology of the networks  via the distances between   persistent diagrams.\end{enumerate}

We point out that the networks that we consider in this paper are very noisy. Metaphorically speaking, what we are trying  to  do here is to quantify the `shape of noise'.

We illustrate our   procedure by investigating financial time series for the US financial crisis of $2007 -  2008$.  The time-varying network that we consider is the cross correlation network $C=(c_{i,j})$ of the stock returns for the companies in the Dow Jones Industrial Index (DJIA); the nodes of the network represent the stocks, and the weights of the edges are given by the   distances  $d_{i,j}=\sqrt{2(1-c_{i,j})}$. Following the process described above, we compute the time series of the distances between the persistent diagrams at time $t$  and the reference persistent diagram at  initial time $t_0$. The conclusion is that these time series display a significant change prior to the critical transition (i.e., the peak of the crisis), which shows that the stock correlation network undergoes significant changes in its topological structure.

For the computation of persistent diagrams and their mutual
 distances we use the \emph{R}  	
package \emph{TDA} \cite{Fasy}.

\section{Background}\label{sec:background}

We provide a brief, largely self-contained, review of the persistent homology method, and describe how to use it to analyze the topology of weighted networks. Some general  references and applications include \cite{Edelsbrunner2008,Nicolau2011,Chazal2012,Adler,BerwaldG14,BerwaldGVJ15,Munch2016,Kramar16}.

\subsection{Persistent homology}
\label{sec:persistent_homology}
Persistent homology is a computational method to extract  topological features from a  given data set (e.g., a point-cloud data set or a weighted network) and rank them according to some threshold parameter (e.g., the distance between data points or the weight of the edges). Topological features that are only visible at low levels of the parameter are ranked lower than topological features that are  visible at both low and high levels. For each value of the threshold parameter one builds a simplicial complex (i.e.,  a space
made from simple pieces -- geometric  simplices, which are identified combinatorially along faces). In our case, the vertices correspond to the data points  and the  simplices are determined by the proximity of data points. When the threshold parameter is varied, the corresponding simplicial complexes form a filtration (i.e., an ordering of the simplicial complexes that is compatible with the ordering of the threshold values). Then one tracks the topological features (e.g., connected components, `holes' of various dimensions) of the simplicial complexes across the filtration, and record for each topological feature the  value of the parameter at which that feature appears for the first time (`birth value'), and the   value of the parameter at which the feature disappears (`death value').
We now give  technical details.

A simplicial complex $K$ is a set of simplices $\{\sigma\}$ of various dimensions that satisfies the following conditions:
(i)~any face of a simplex   $\sigma\in K$ is also in $K$, and (ii) the intersection of any two simplices $\sigma _{1}, \sigma _{2}\in K$ is either $\emptyset$  or a face of both $\sigma _{1}$  and $\sigma _{2}$.

Given a simplicial complex $K$, we denote by  $H_i(K)$  the $i$-th  homology group  with coefficients in $\mathbb{Z}_2$. This is a free abelian group whose generators  consists of certain chains of $i$-dimensional simplices (i.e., cycles that are not boundaries).
Note that $H_i(K)=0$ for  $i\geq m+1$. The  generators of the $i$-th homology group account for the `independent holes' in $K$ at dimension $i$. For example,   the number of $0$-dimensional generators equals   that of connected components of $K$, the number of $1$-dimensional generators  equals that  of `tunnels' (or `loops'), the number of $2$-dimensional generators  equals that of `cavities', etc. For a reference, see, e.g., \cite{kaczynski2004computational}.

A \emph{filtration} of $K$ is a mapping  $a\in A\mapsto \mathscr{F}(a):= K_a\subseteq K$, from a  (totally ordered) set of parameter values $A\subseteq\mathbb{R}$ to a set of simplicial sub-complexes of $K$, satisfying the \emph{filtration condition}:
$a\leq a'\,\Rightarrow K_a\subseteq K_{a'}$.
For any filtration of simplicial complexes
$a \mapsto  K_a $ the corresponding homology groups also form a filtration $a \mapsto  H_i(K_a)$, that is,   $a\leq a'\,\Rightarrow H_i(K_a)\subseteq H_i(K_{a'})$.

For  $a\leq a'$,  the inclusion   $H_i(K_a)\subseteq H_i(K_{a'})$   induces a group homomorphism  $h^{a,{a'}}_i:H_i(K_a)\to H_i(K_{a'})$, in all $i$.
Let  $H_i^{a,{a'}}=\textrm{Im}(h^{a,{a'}}_i)$ be the image of $h^{a,{a'}}_i$ in $H_i(K_{a'})$.
We say that a  homology class $\gamma\in H_i(K_{b})$ is \emph{born} at the parameter value $a=b$ if $\gamma\not \in H_i^{{b}-\delta,{b}}$ for any $\delta>0$. If $\gamma$ is born at $K_{b}$ then we say that it \emph{dies} at the parameter value $a=d$, with ${b}\leq d$,  if $\gamma$ coalesces  with an older class in $H_i(K_{d-\eps})$ as we go from $K_{d-\eps}$ to $K_{d}$ for $\eps>0$, that is, $h^{{b},d-\eps}(\gamma)\not\in H_i^{{b}-\delta,d-\eps}$ for any small $\eps,\delta>0$, but $h^{{b},d}_i(\gamma)\in H_i^{{b}-\delta,d}$ for some small $\delta>0$. If $\gamma$ is born at $K_{d}$ but never
dies then we say that it dies at infinity. Thus, we have a value
$b(\gamma)=b$ and a death value $d(\gamma)=d$ for each generator $\gamma$
that appears in the filtration  of homology groups.
The persistence, or `life span'  of the class $\gamma$ is the difference between the two values, $\textrm{pers}(\gamma) = d(\gamma)-b(\gamma)$.

The $i$-th persistent diagram of the filtration $\mathscr{F}$  is defined as a multiset  ${P}_i$ in $\mathbb{R}^2$, for $i=0,\ldots,m$, obtained as follows:
\begin{itemize} \item For each class $\gamma_i$ we assign a point $z_i=(b_i,d_i)\in\mathbb{R}^2$ together with a multiplicity $\mu_i(b_i,d_i)$;  where $b_i$ is the parameter  value   when $\gamma_i$ is born, and $d_i$ is  the parameter  value  when  $\gamma_i$ dies.
The multiplicity
$\mu_i(b_i,d_i)$ of the point $z_i=(b_i,d_i)$ equals the number of classes $\gamma_i$ that are born at ${b_i}$ and die at ${d_i}$.
This multiplicity is finite since the simplicial complex is finite.
\item In addition, $P_i$ contains all points in the positive diagonal  of $\mathbb{R}^2$. These points represent  all trivial homology generators that are born and die at every level. Each point on the diagonal has infinite multiplicity.
\item The axes of the persistent diagram are birth values on the horizontal axis and death values on the vertical axis.
\end{itemize}

\begin{figure}\centering
\includegraphics[width=1\textwidth]{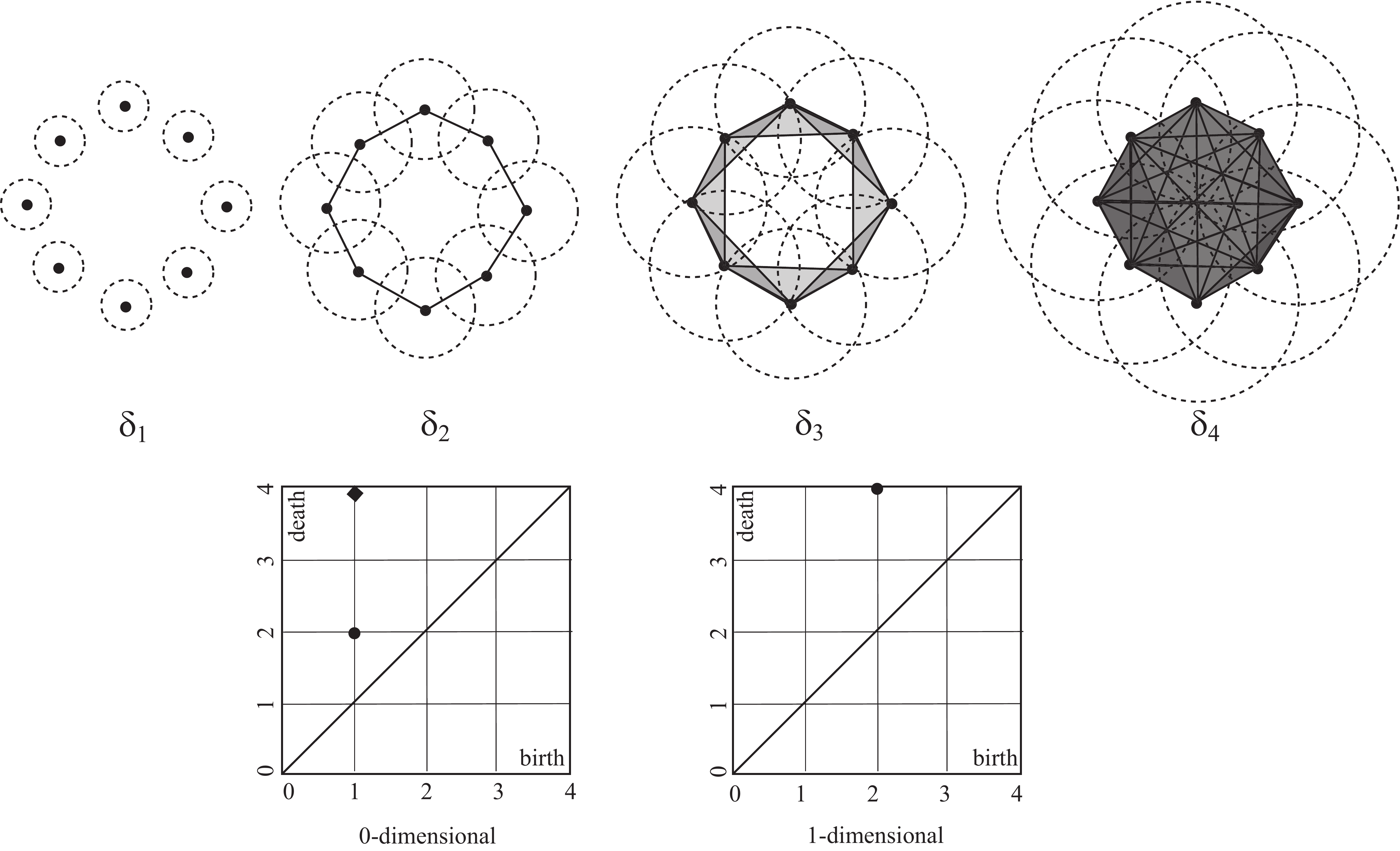}
\caption[]{A point-cloud data set representing a `noisy'
  circle,  together with a filtration of simplicial complexes, corresponding to some threshold parameter values
  $\delta_1<\delta_2<\delta_3<\delta_4$.   The 0-dimensional and the 1-dimensional persistence diagrams are shown at the bottom of the figure. At
  $\delta_1$ there are $8$ connected components and no $1$-dimensional hole. At
    $\delta_2$ the $8$ connected components coalesce into a single one, indicated by the point $(1,2)$  in the $0$-dimensional diagram, which has multiplicity $7$;
  also, a $1$-dimensional hole is born.  There is no topological change at
  $\delta_3$. At  $\delta_4$ the $1$-dimensional hole gets filled in and dies, indicated by the
  the point $(2,4)$ in the $1$-dimensional diagram;    the single connected
  component continues living  for ever, and is represented by
  $\blacklozenge$.
}
    \label{persistence}
\end{figure}

An  illustration of persistent diagrams for a simple example of point-cloud data set    and a filtration of simplicial complexes   is shown in Fig. \ref{persistence}.

The space of persistent diagrams can be endowed with a metric space structure.
A standard metric that can be used is  the \emph{degree $p$ Wasserstein distance} (earth mover distance), with $p>0$. This is defined by
$\displaystyle D_{p}( P^1_i, P^2_i)=  \inf_{\phi}\left[\sum_{q\in P^1_i} \|q- \phi(q)\|^p_\infty\right]^{1/p},$
where the summation is  over all bijections $\phi: P^1_i\to P^2_i$.   When $p=\infty$ the Wasserstein distance $D_\infty$  is known as the
bottleneck distance.
Since the diagonal set is by default part of all persistent diagrams, the pairing of points between $ P^1_i$ and $ P^2_i$ via $\phi$ can include pairings between off-diagonal points and diagonal points.

We note that different value of the degree $p$ yield different types of  measurements of the distances between persistent diagrams. Using  $p=\infty$, the corresponding distance $D_\infty$ only measures the distance between the most significant features (farthest from the diagonal)  in the diagrams,  matched via some appropriate $\phi$. Using $p\geq 1$  large, the corresponding distance $D_p$ puts more weight on the significant features (farther from the diagonal) than on the least significant ones (closer to the diagonal). Using $p>0$ small has just the opposite effect on the measurement.

One of the remarkable properties of persistent diagrams is their
robustness, meaning that small changes in the initial   data
produce persistent diagrams that are close to one another relative to
Wasserstein metric.  The essence of the stability
results  is that the
persistent diagrams depend Lipschitz-continuously on
data. For details see~\cite{Chazal2009,Cohen-Steiner2010,EdelsbrunnerM2012}.

\subsection{Persistent homology of weighted networks}
A weighted network   is a pair consisting of a graph $G=G(V,E)$ and a weight function associated to its edges $w:E\to[0,+\infty)$; let $\theta_{\textrm{\textrm{max}}}=\max(w)$. In the sequel we will only consider graphs that are simple  and undirected. In examples, the weight function is chosen so that nodes with similar characteristics are linked together.

One standard recipe to investigate the topology  of weighted graphs is via thresholding, that is, by considering only those edges whose weights are below (or above) some suitable threshold, and study the features of the resulting graph.
Of course, the choice of the threshold value makes a difference in the topology of the resulting graph. Using persistent homology, we can  extract the topological features for each threshold graph, and represent all these features, ranked according to their `life span', in a persistent diagram. We now give technical details.

For each $\theta\in[0,\theta_{\textrm{max}}]$, we consider the sub-level sets of the weight function, that is, we restrict to subgraphs $G(\theta)$ which keep  all edges of weights $w$ below or equal to the threshold $\theta$.
The graphs obtained by restricting to successive thresholds have the  filtration property, i.e.,  $\theta\leq \theta' \Longrightarrow
 G(\theta)\subseteq G(\theta')$.
In a similar way, we can consider super-level sets, by  restricting to subgraphs $G(\theta)$ which keeps all edges of weights $w$ above or equal to the threshold $\theta$. Super-level sets can be thought of as sub-level sets of the weight function $w'=\theta_{\textrm{max}}-w$.

For each threshold graph $G(\theta)$ we  construct the \emph{Rips complex} (clique  complex) $K=X(G(\theta))$. This is defined as   the simplicial complex with all complete subgraphs (cliques) of $G(\theta)$ as its faces.
That is,  the $0$-skeleton
of $K$   consists of just the vertices of $G(\theta)$, the $1$-skeleton of all vertices and edges -- which is the graph $G(\theta)$ itself --, the $2$-skeleton of all vertices,
edges, and filled triangles, etc.   High dimensional cliques correspond to  highly interconnected  clusters of nodes  with  similar characteristics (as encoded by the weight function). The filtration of the threshold subgraphs yields a corresponding filtration of the Rips complexes $\theta\mapsto K_\theta:=X(G(\theta))$; thus,  $\theta\leq \theta'\Longrightarrow
 K_{\theta}\subseteq K_{\theta'}$.
As it was noted before, the homology groups associated to this filtration satisfy themselves the  filtration property, i.e., $\theta\leq \theta'\Longrightarrow
H_i(K_{\theta})\subseteq H_i(K_{\theta'})$.
From this point on, we can compute the persistent homology and the persistent  diagrams  associated to this filtration, in the manner described in Section~\ref{sec:persistent_homology}.

In Section \ref{sec:correlation_networks} we will only compute persistent diagrams of dimension $0$ and $1$, so we explain in detail the significance of these diagrams in terms of the threshold networks.

A point $(\theta_b,\theta_d)$ in a $0$-dimensional persistent diagram has the following meaning:
\begin{itemize}
\item At the threshold value $\theta_b$ a connected component is born, where each pair of nodes in the component is connected via a path of edges of weights $\theta\leq \theta_b$;
\item  At the threshold value $\theta_d$ two or more connected components coalesce into a single one, via the addition of one or several edges of weight $\theta=\theta_d$ to the threshold network.
\end{itemize}

A point $(\theta_b,\theta_d)$ in a $1$-dimensional persistent diagram has the following meaning:
\begin{itemize}
\item At the threshold value $\theta_b$ a loop of 4 or more nodes is born, whose  nodes are connected in circular order by  edges of weights $\theta\leq \theta_b$; note that a loop of $3$ nodes yields a complete sub-graph in the Rips complex (i.e., a filled triangle),  which carries no $1$-dimensional homology;
\item  At the threshold value $\theta=\theta_d$ one or more loops get covered by filled triangles, due to adding one or more edges of weight $\theta=\theta_d$, thus making the corresponding $1$-dimensional homology generators   vanish.
\end{itemize}

We note that applications of persistent homology to networks also appear, e.g.,  in \cite{Carstens,Horak}.

\section{Detection of critical transitions from correlation networks}
\label{sec:correlation_networks}

\subsection{Correlation networks as weighted networks}
\label{sec:correlation_networks_stocks}
The network that we analyze here is derived from  the DJIA stocks listed as of February 19, 2008. We utilize the time series of the  daily returns  based on the adjusted closing prices $S_i(t)$, i.e.,  $x_i(t)=\frac{S_i(t+\Delta t)-S_i(t)}{S_i(t)},$
where $\Delta t=1$ day, and  the indices $i$ correspond to the individual stocks.
We restrict to the data from January 2004 to September 2008 (when Lehman Brothers filed for bankruptcy).

Now we define the weighted network $G(V,E)$ that we analyze. The  vertices $V$ of the network  correspond  to the individual  DJIA stocks. Each pair of distinct vertices $i,j\in V$  is connected by an edge $e$, and each edge is assigned a  weight $w(e,t)$ at time $t$ defined as follows:
\begin{itemize}
\item Compute the Pearson correlation coefficient $c_{i,j}(t)$ between the nodes $i$ and $j$  at time $t$, over a time horizon $T$,  by
    \[c_{i,j}(t)=\frac{\sum_{\tau=t-T}^{t}(x_i(\tau)-\bar x_i)(x_j(\tau)-\bar x_j)}{\sqrt{\sum_{\tau=t-T}^{t}(x_i(\tau)-\bar x_i)^2}\sqrt{\sum_{\tau'=t-T}^{t}(x_j(\tau')-\bar x_j)^2}},\]
    where $\bar x_i$, $\bar x_j$ denote the averages of $x_i(t)$, $x_j(t)$ respectively, over the time interval $[t-T,t]$;
    \item Compute the  distance between the nodes $i$ and $j$, $d_{i,j}(t)=\sqrt{2(1-c_{i,j}(t))}$  -- the fact that the metric axioms are satisfied follows easily from the properties of correlation;
    \item Assign the weight $w(e,t)=d_{i,j}(t)$ to  the edge $e$ between $i$ and $j$.
\end{itemize}

For the computation of the correlation via the  Pearson estimator,  there is empirical evidence against using longer time horizons  when non-stationary behavior is present. Therefore, in our computation we use a  rather short time horizon  of $T=15$;  we also use the arithmetic return rather than the standard log return. For an argument  in support of these choices see \cite{Munix}.

The range of values of $d_{i,j}$ is $[0,2]$. Note that $d(i,j)=0$ if the nodes  $i$ and $j$ are perfectly correlated, and $d(i,j)=2$ if the nodes are perfectly anti-correlated. Edges between  correlated nodes have smaller weights, and edges between uncorrelated/anti-correlated nodes have bigger weights. Since correlation between stocks is of interest,  we focus on edges with low values of $d$.

In the sequel, we will consider both sub-levels sets and super-level sets of the weight function.

Each  sub-level set  of the weight function $w$, at a threshold level $\theta\in[0,2]$, yields a subgraph $G(\theta)$ containing only those edges for which $0\leq d_{i,j}\leq\theta$,
that is, $\displaystyle G(\theta)=\{e=e(i,j)\,|\, 1-\frac{1}{2}\theta^2\leq c_{i,j}\leq 1.\}$
When $\theta$ is small, $G(\theta)$ contains only edges between highly-correlated nodes.
As $\theta$ is increased up to the critical value $\sqrt{2}= 1.414214$ edges between low-correlated nodes are progressively added to the network.
As $\theta$ is increased further, edges between  anti-correlated  nodes appear in  the network.

Each  super-level set of the weight function $w=d$ can be conceived as a sub-level sets for the dual weight function $w'=2-d$. The sub-level set $G(\theta)$ for this weight-function contains only those edges for which $d_{i,j}\geq 2-\theta$, hence $\displaystyle G_{w'}(\theta)=\{e=e(i,j)\,|\, -1\leq c_{i,j}\leq 1-\frac{1}{2}(2-\theta)^2.\}$ When $\theta$ is small, $G_{w'}(\theta)$ contains only edges between anti-correlated nodes. When  $\theta$ crosses the critical value $2-\sqrt{2}= 0.5857864$,  edges between low-correlated nodes are progressively added to the network. As $\theta$ is increased further towards the highest possible value of  $2$, highly-correlated nodes are  added to the network.

Sub-level sets and super-level sets produce very different type of networks, and they furnish complementary information. We will discuss this in Section \ref{sec:pers_correlation}.

\subsection{Persistent diagrams of correlation networks}
\label{sec:pers_correlation}

\begin{figure}
\centering\[\begin{array}{ccc}
\includegraphics[width=0.28\textwidth]{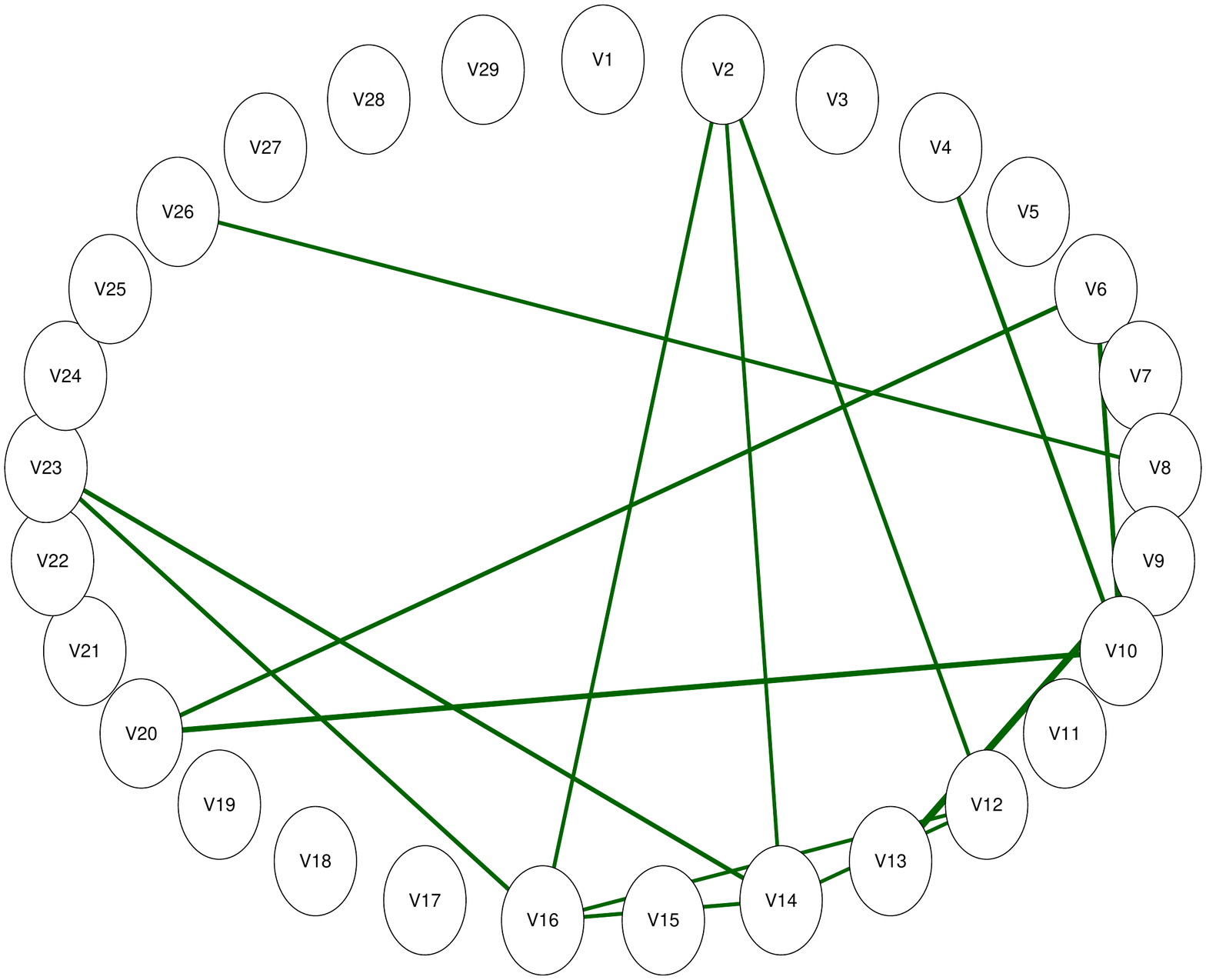}&
\includegraphics[width=0.28\textwidth]{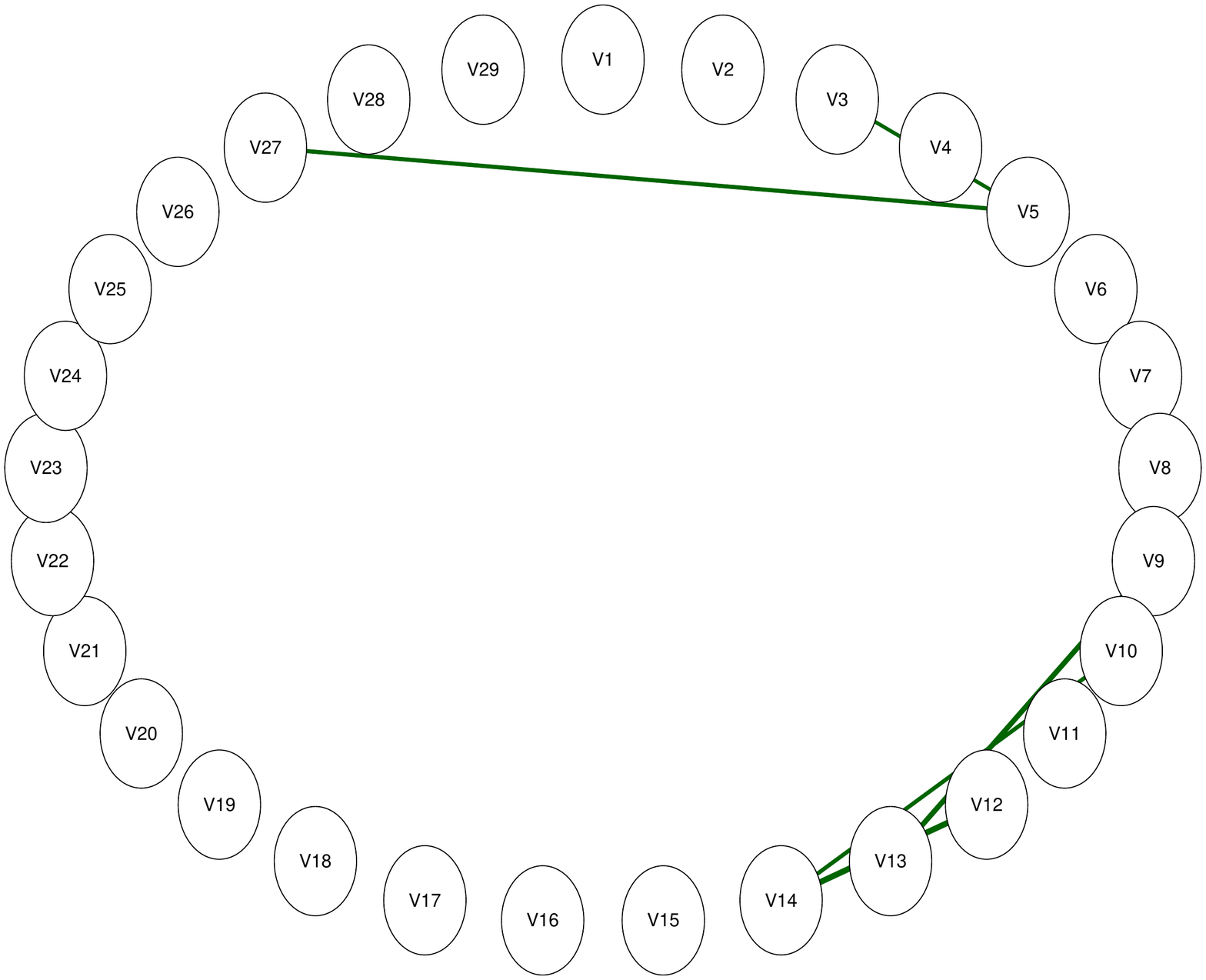}&
\includegraphics[width=0.28\textwidth]{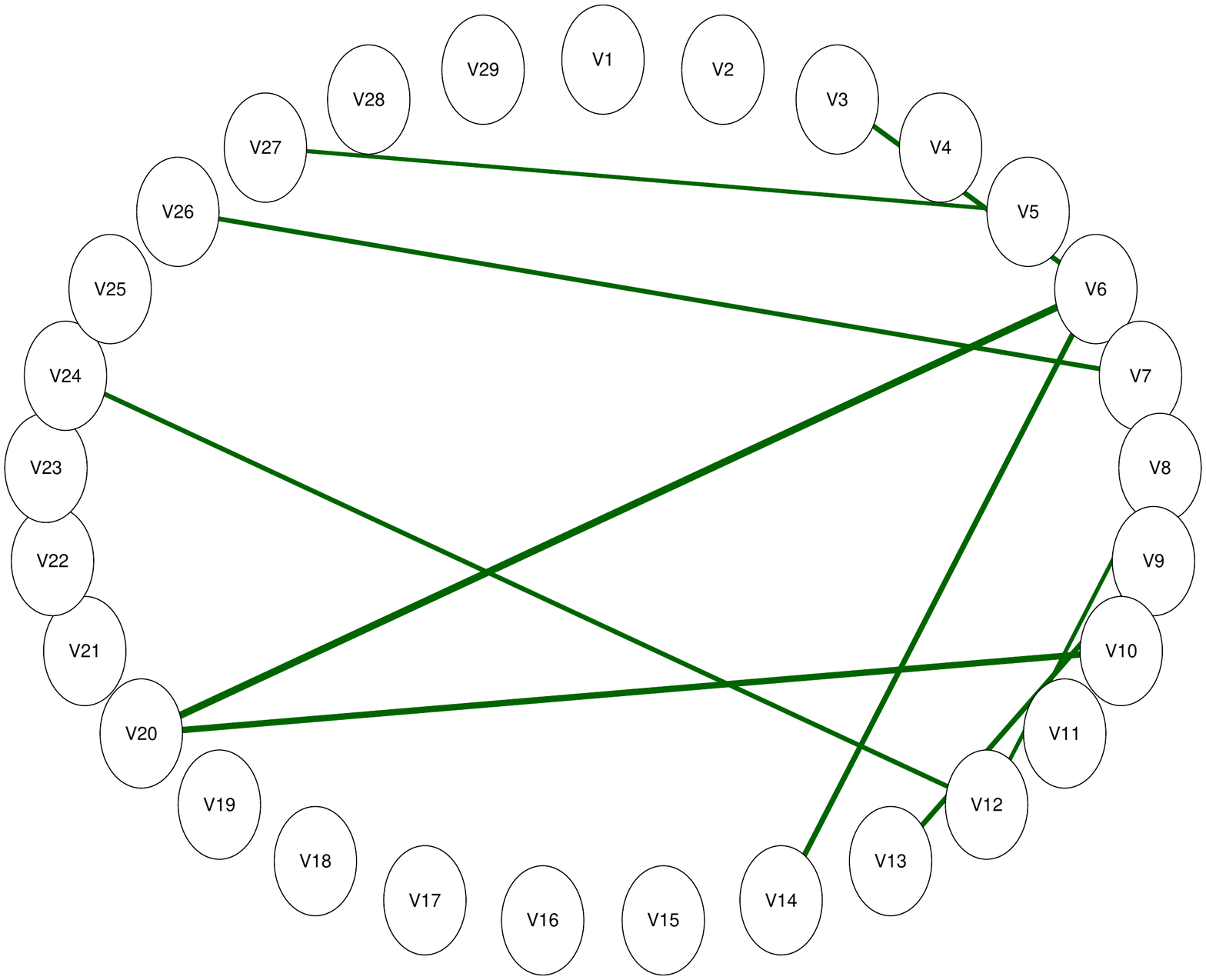}\\
\includegraphics[width=0.28\textwidth]{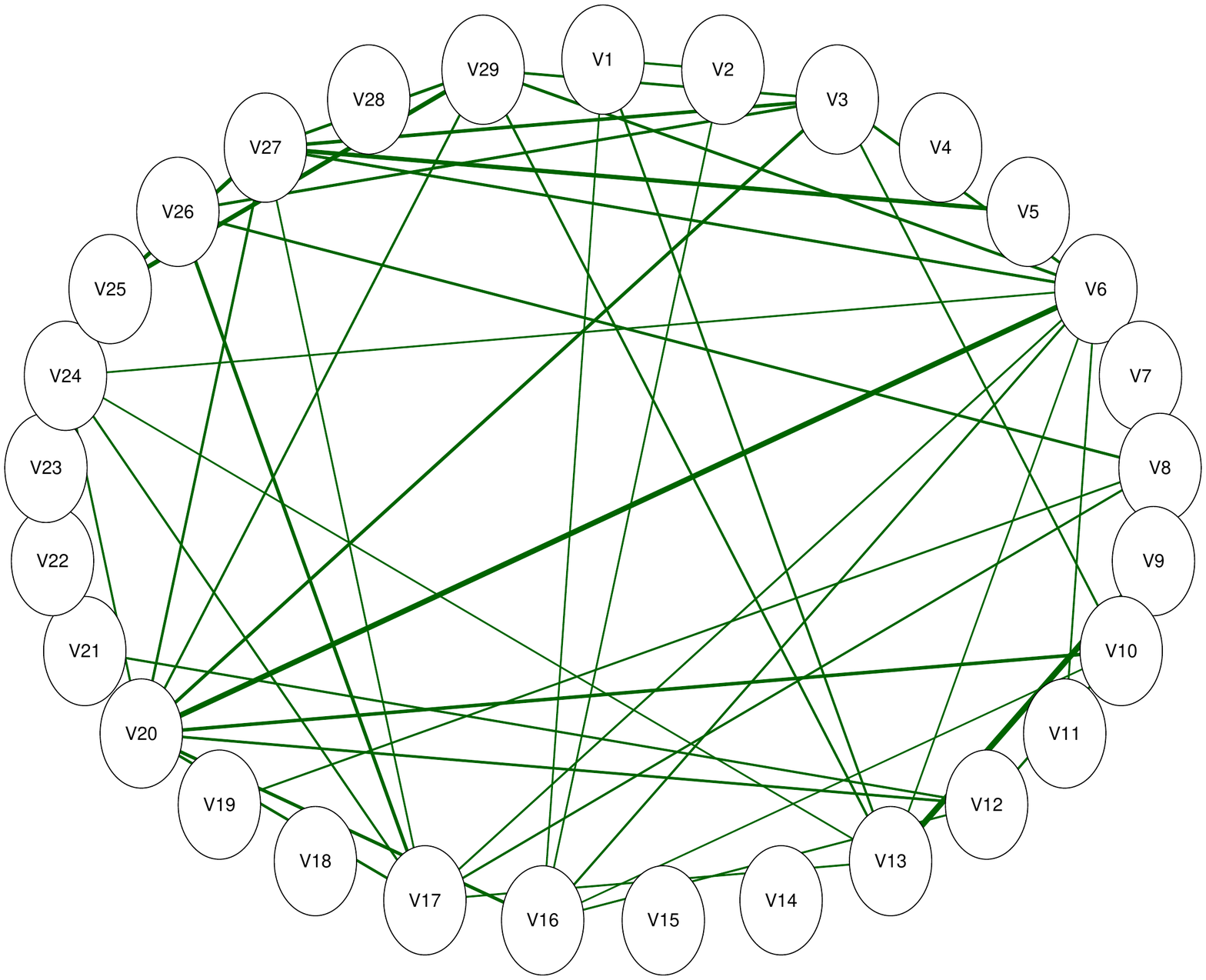}&
\includegraphics[width=0.28\textwidth]{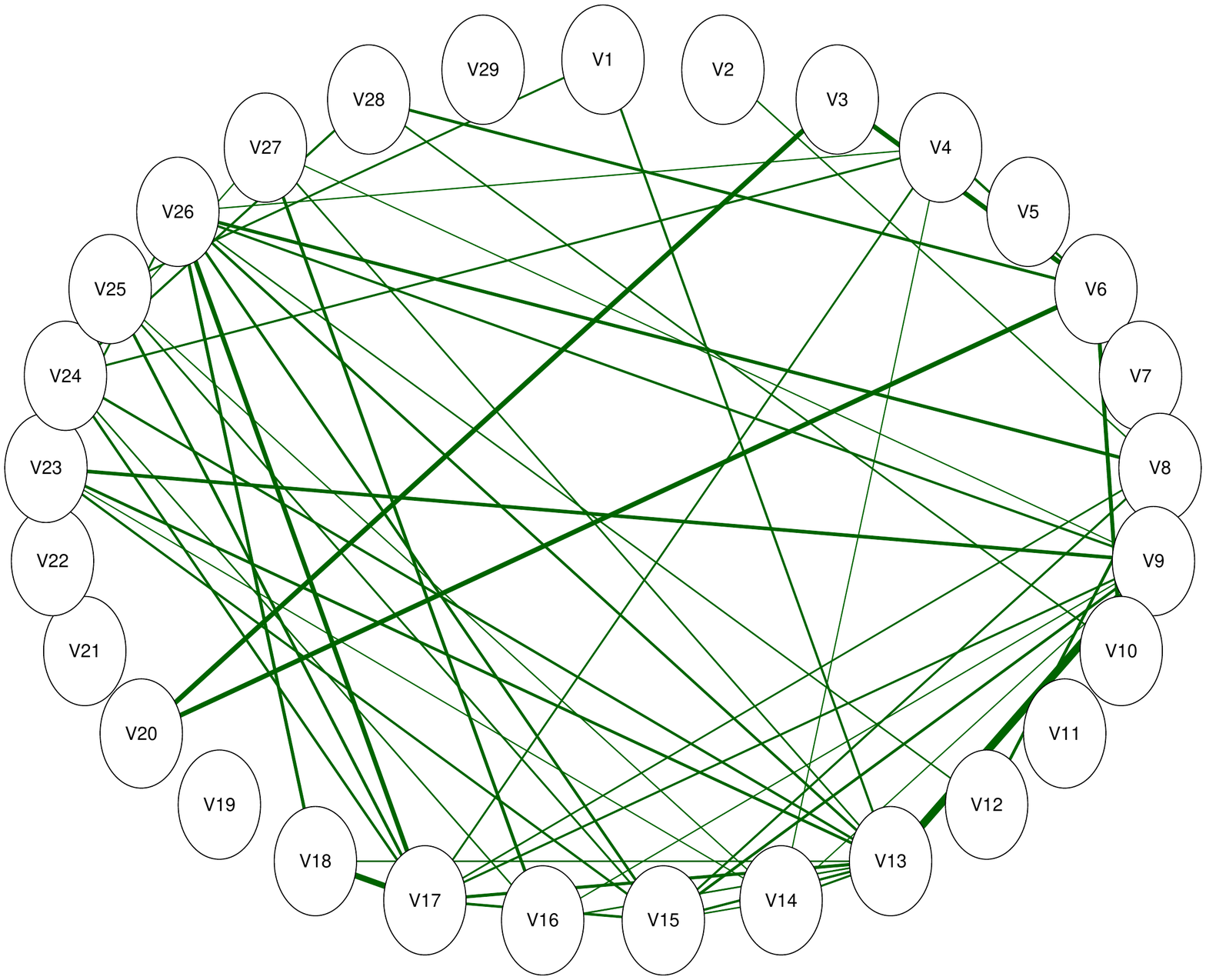}&
\includegraphics[width=0.28\textwidth]{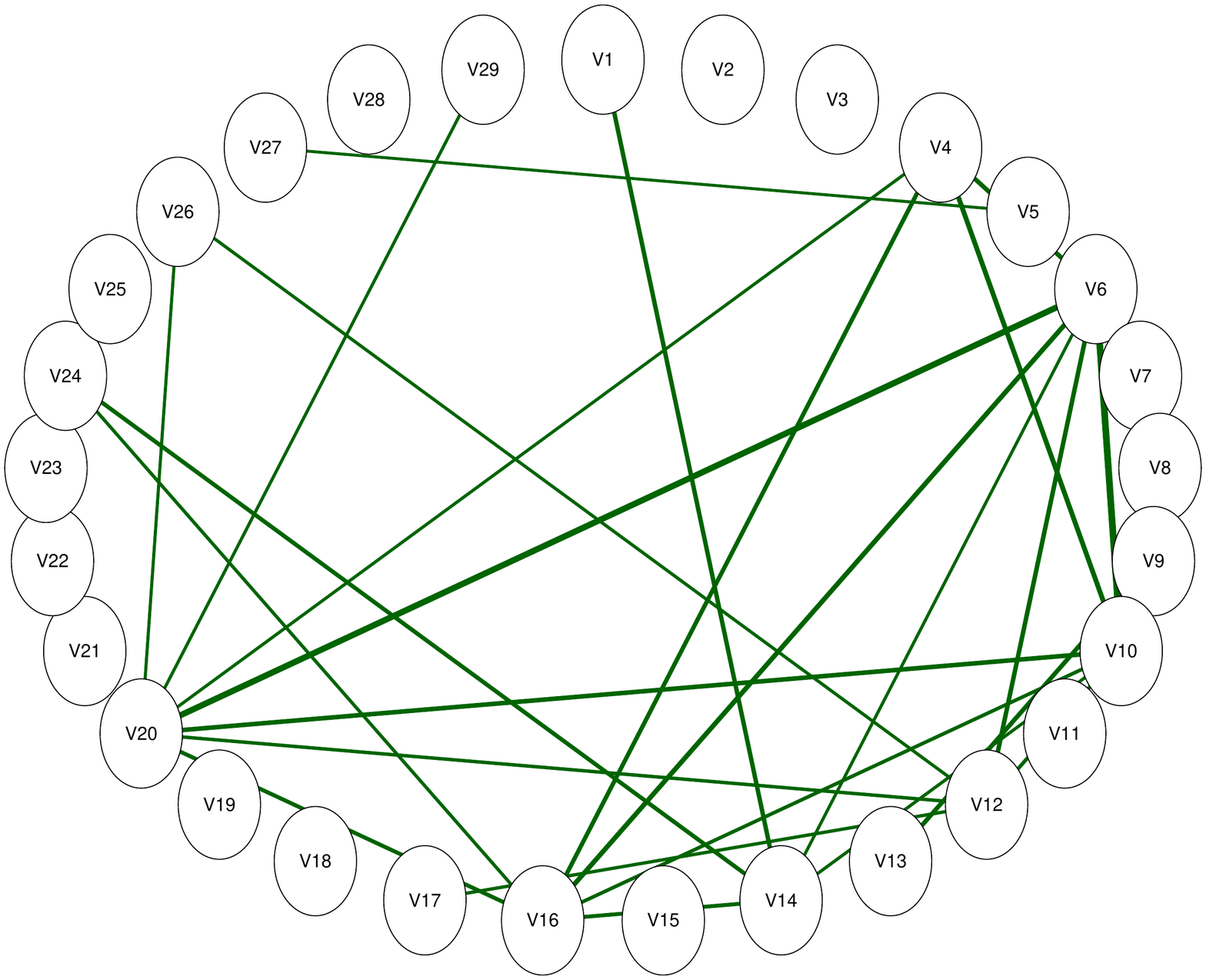}
\end{array}\]
\caption{Threshold correlation networks}
\label{fig:networks}
\end{figure}

In this section we use persistent homology to quantify the changes in the topology of the correlation networks when approaching a critical transition. For illustrative purposes, we show some correlation networks  in Figure \ref{fig:networks}; the top three networks represent instants of time far from the the beginning  of the 2007-2008 financial crisis, while the bottom three diagrams represent instants of time  preceding the crisis.

We compute  persistent homology in dimensions $0$ and $1$ for the correlation network from Section~\ref{sec:correlation_networks_stocks}. We do not consider  higher-dimensional persistent homology because  the network is very small, so the presence higher-dimensional cliques is likely accidental.

\begin{figure}
\centering\[\begin{array}{ccc}
\includegraphics[width=0.3\textwidth]{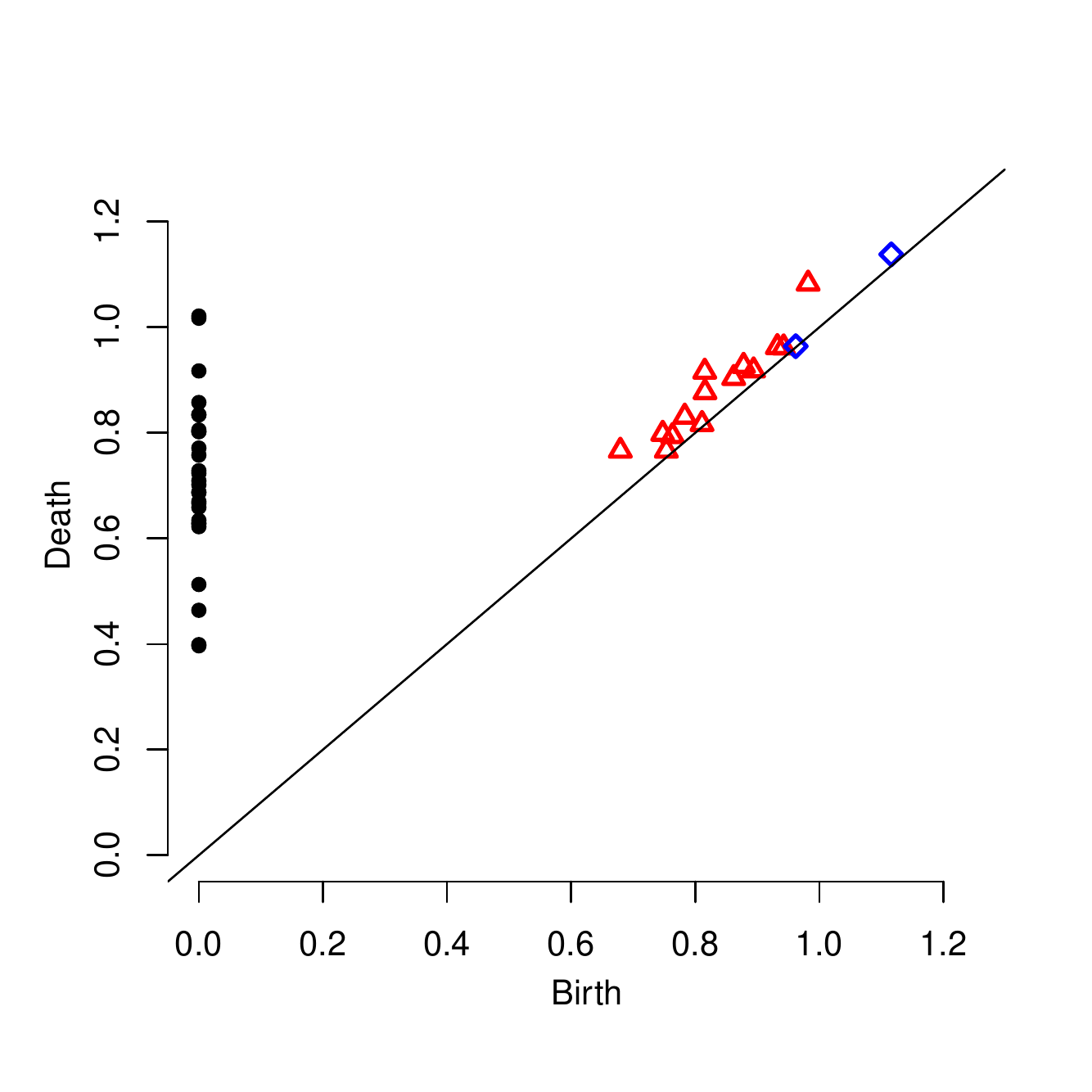}&
\includegraphics[width=0.3\textwidth]{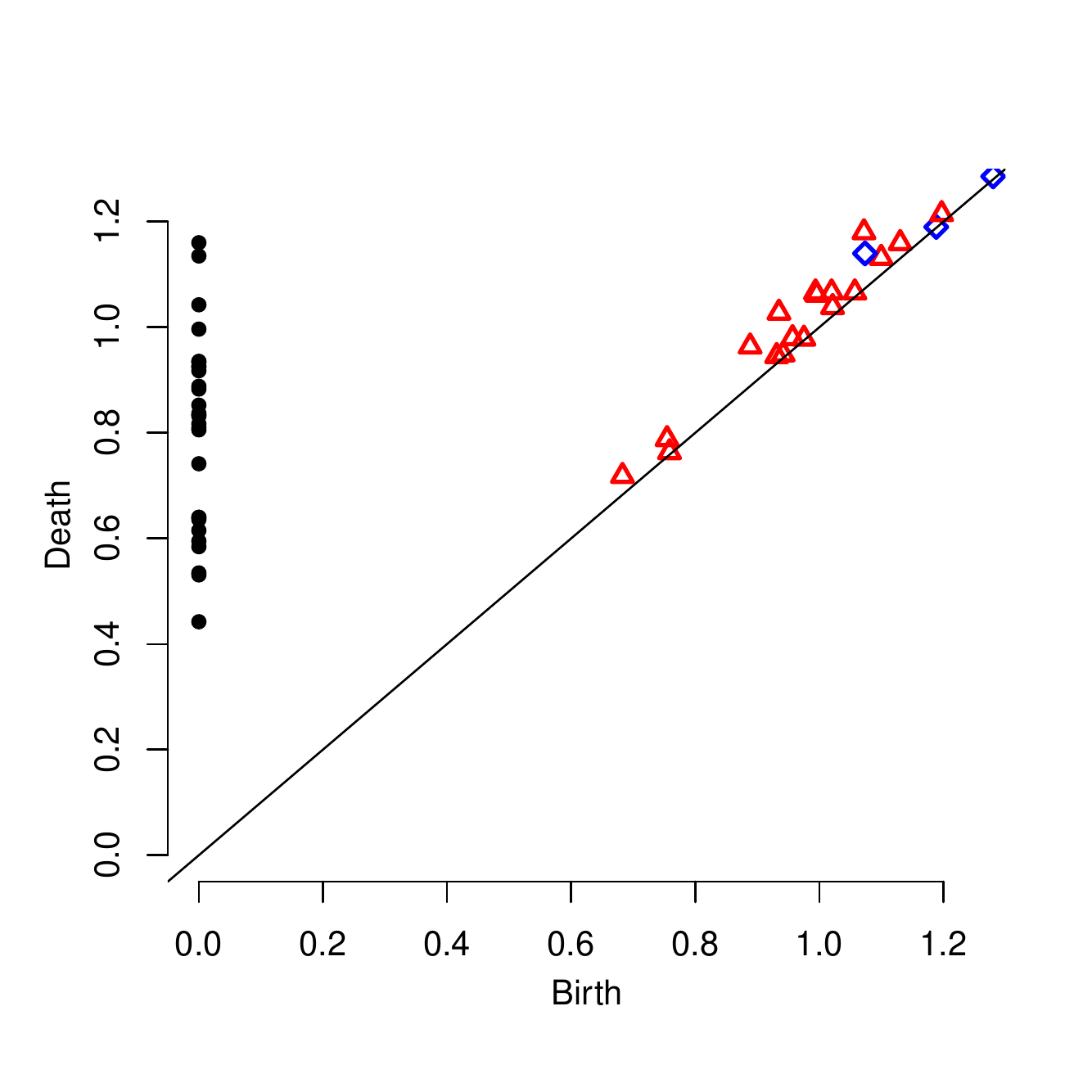}&
\includegraphics[width=0.3\textwidth]{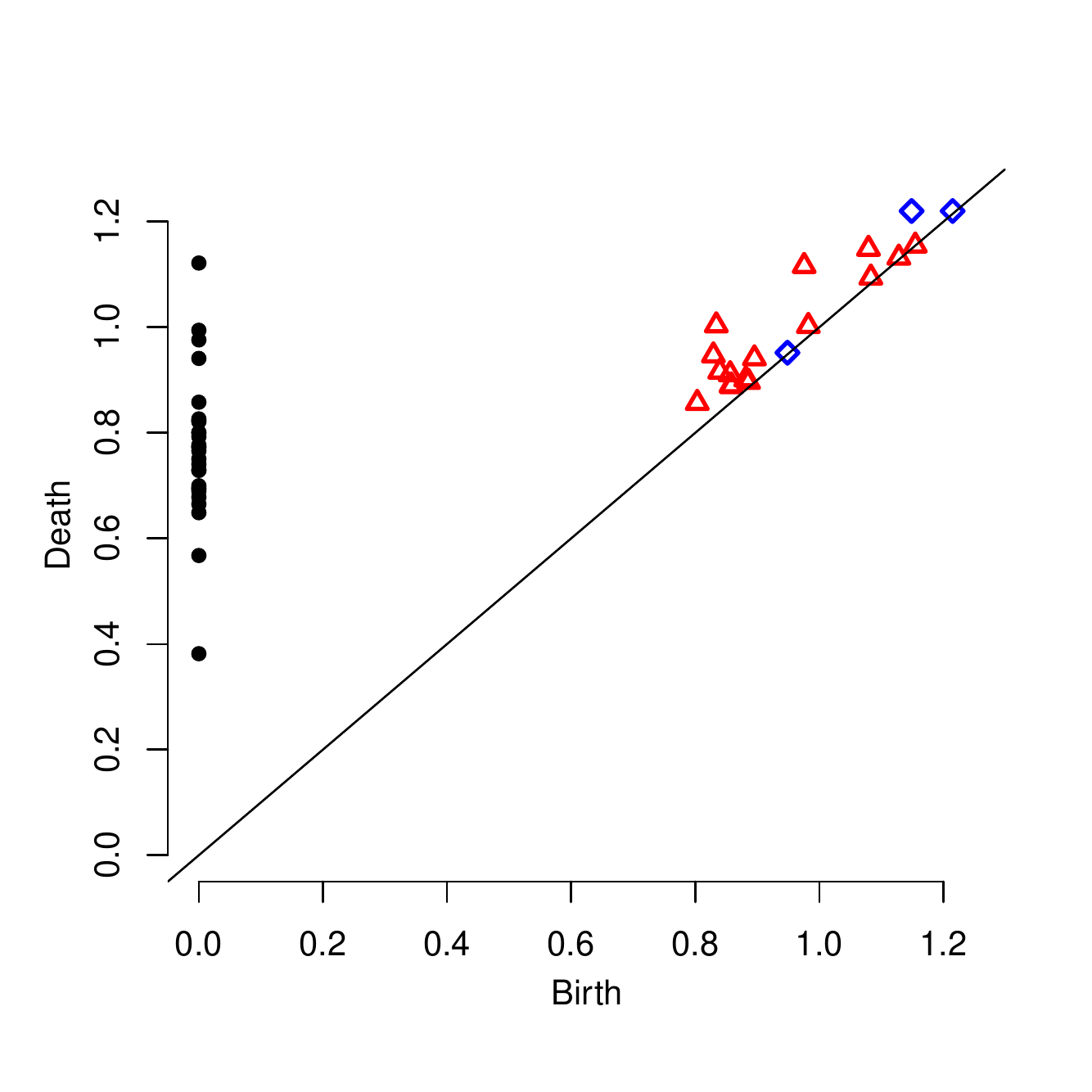}\\
\includegraphics[width=0.3\textwidth]{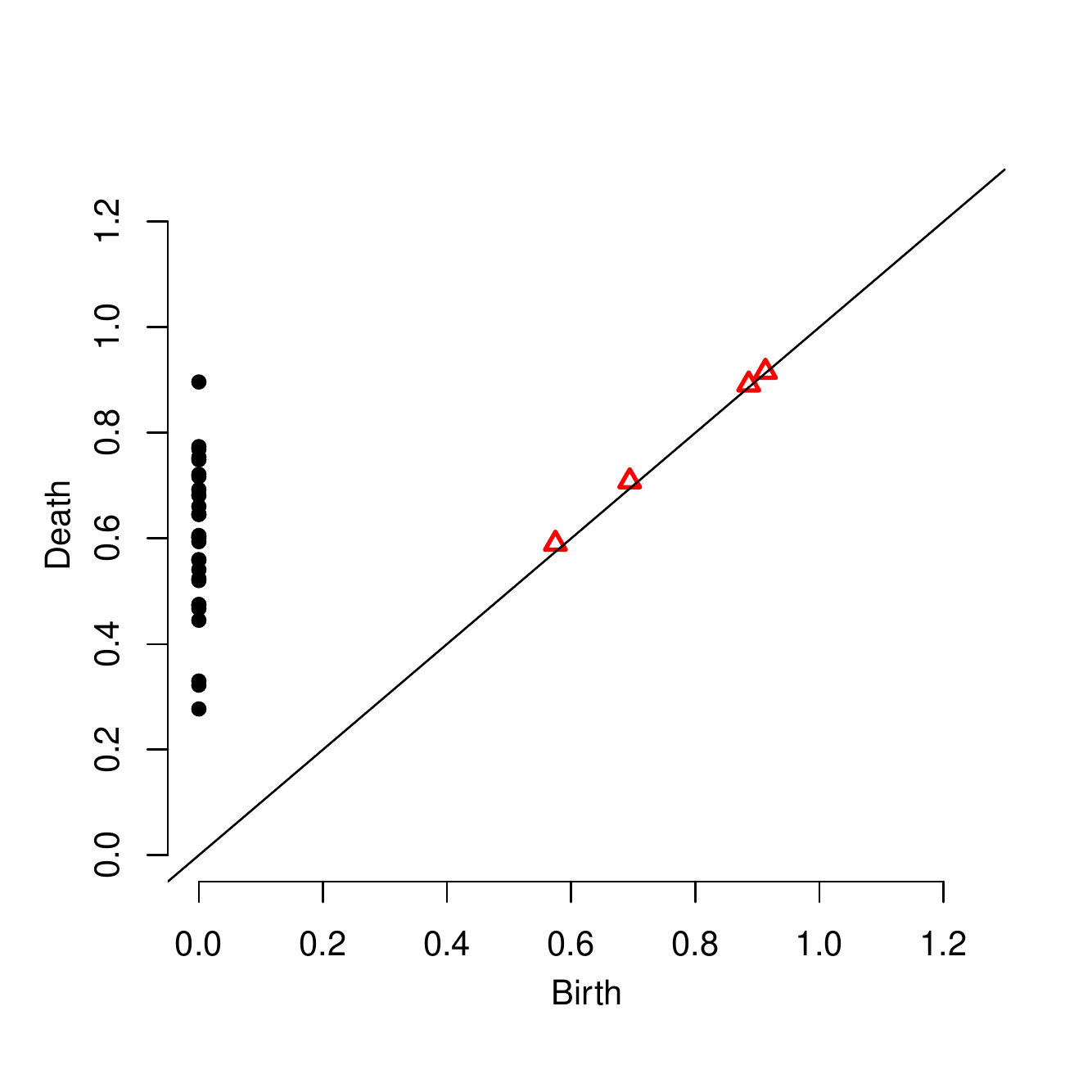}&
\includegraphics[width=0.3\textwidth]{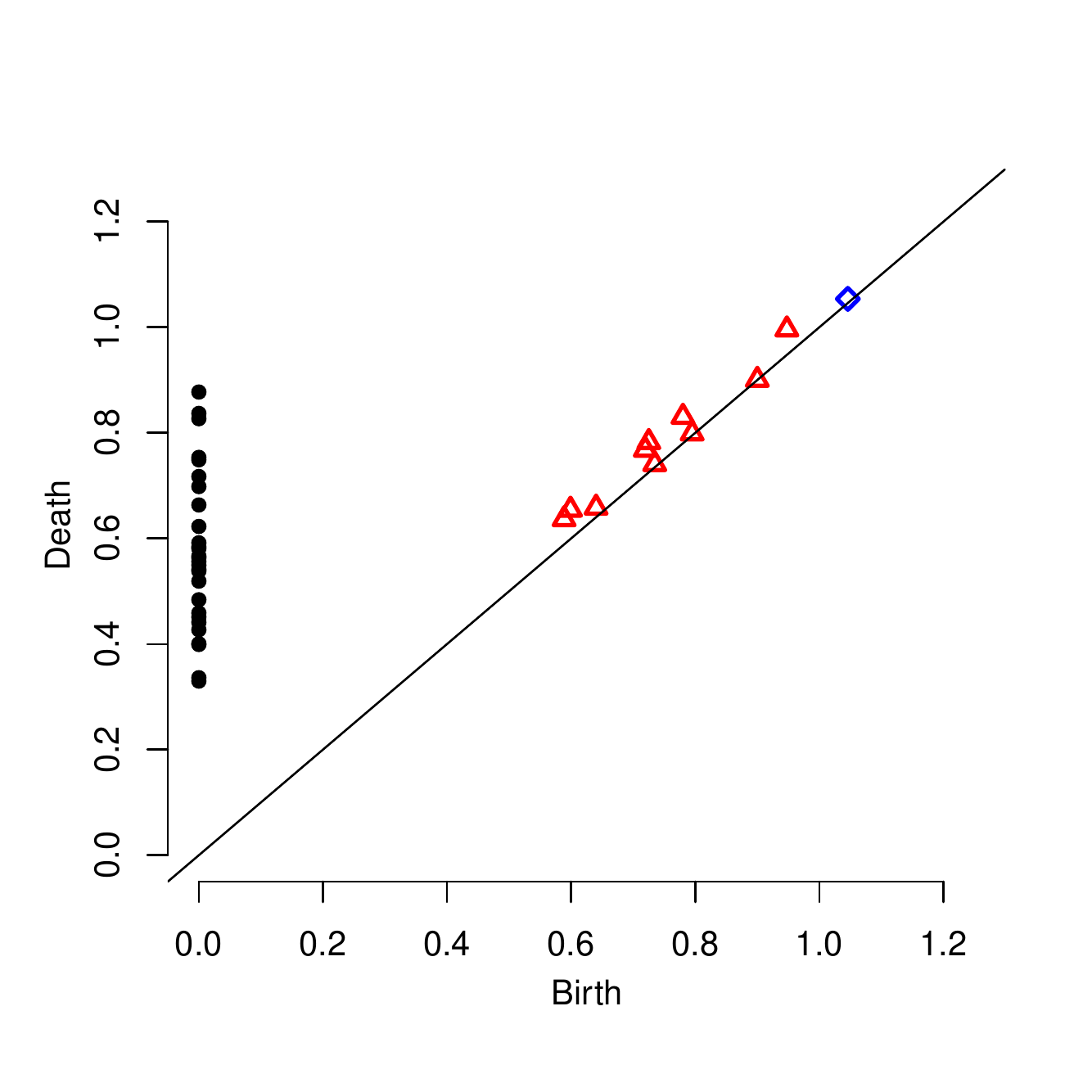}&
\includegraphics[width=0.3\textwidth]{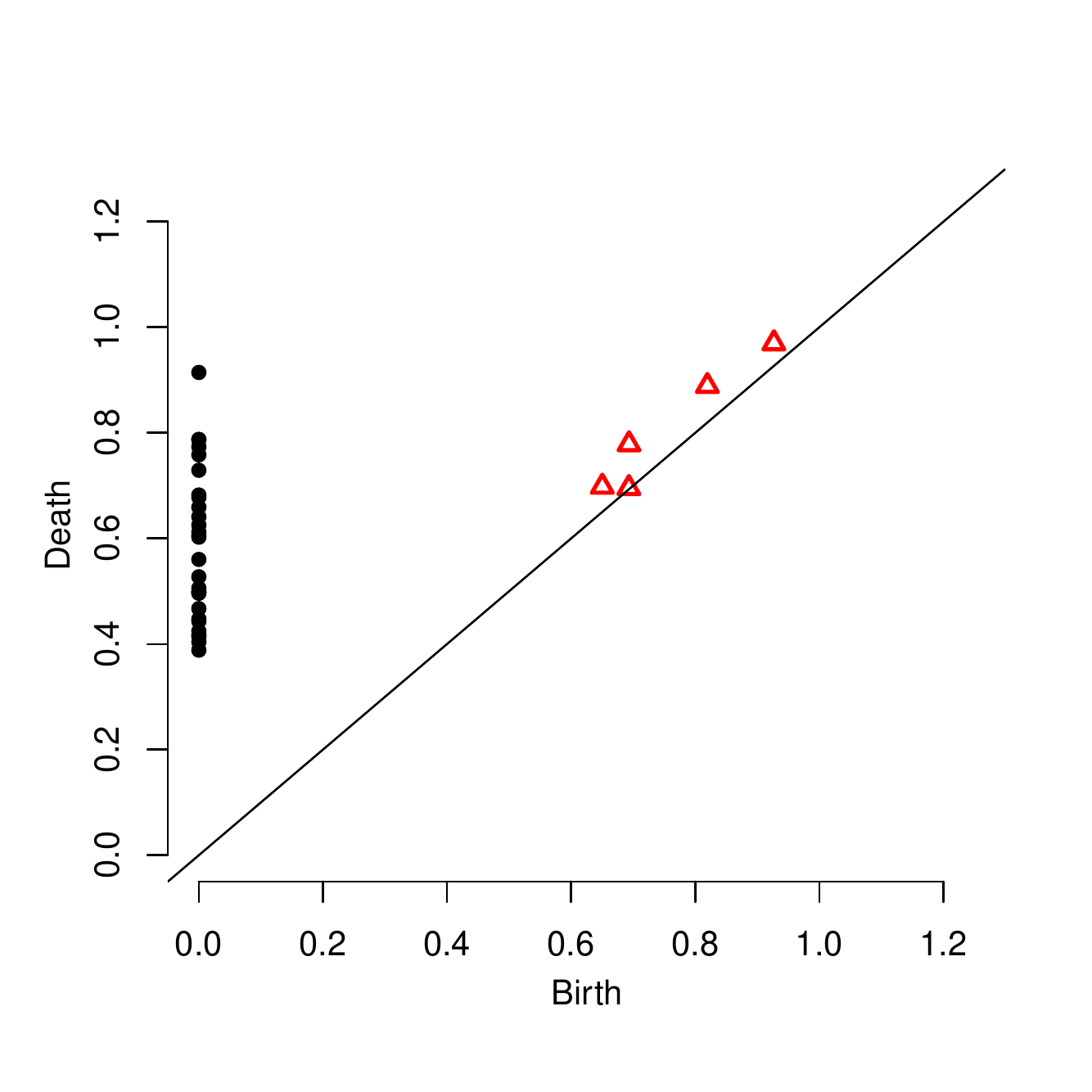}
\end{array}\]
\caption{Persistent diagrams  (sub-level sets)}
\label{fig:sublevel-diagrams}
\end{figure}

First, we consider threshold networks given by the sublevel sets of the weight function $w=d$.
Several persistent  diagrams are shown in  Fig.~\ref{fig:sublevel-diagrams}. The top three diagrams correspond to instants of time far from the the beginning  of the 2007-2008 financial crisis, while the bottom three diagrams correspond to instants of time  preceding the crisis.

The $0$-dimensional persistent homology provides information on how the network connectivity changes as the value of $\theta$ in increased from $0$ to $2$. Each black dot on the persistent diagram corresponds to one (or several) connected component of the graph. The horizontal coordinate of each dot is $0$, since all components are born at threshold value $\theta=0$. The vertical coordinate of a dot gives  the threshold value  $\theta$ at which a connected component dies, by joining together with another connected component. The dot with highest vertical coordinate (other than  $2$) gives the threshold value $\theta$ for which the graph becomes fully connected. A dot at $2$ (the maximum value) indicates that once the graph is fully connected, it remains fully connected (hence the component never dies) as $\theta$ is further increased. Dots with lower vertical coordinates indicate threshold values for which smaller connected components consisting of highly correlated nodes  die, i.e., coalesce together into larger components.  Dots with higher  vertical coordinates correspond to death of connected components due to the appearance of edges between uncorrelated or anti-correlated nodes. We recall that the critical value of $\theta$ that marks the passage from correlation to anti-correlation is $1.41$.
Inspecting the diagrams in  Fig.~\ref{fig:sublevel-diagrams} we see a concentration of dots with higher vertical coordinates in the first period, and a
a concentration of dots with lower vertical coordinates in the second period.
There is less correlation in the network in the first period than in the second period.

\begin{figure}
$\begin{array}{cc}
\includegraphics[width=0.5\textwidth]{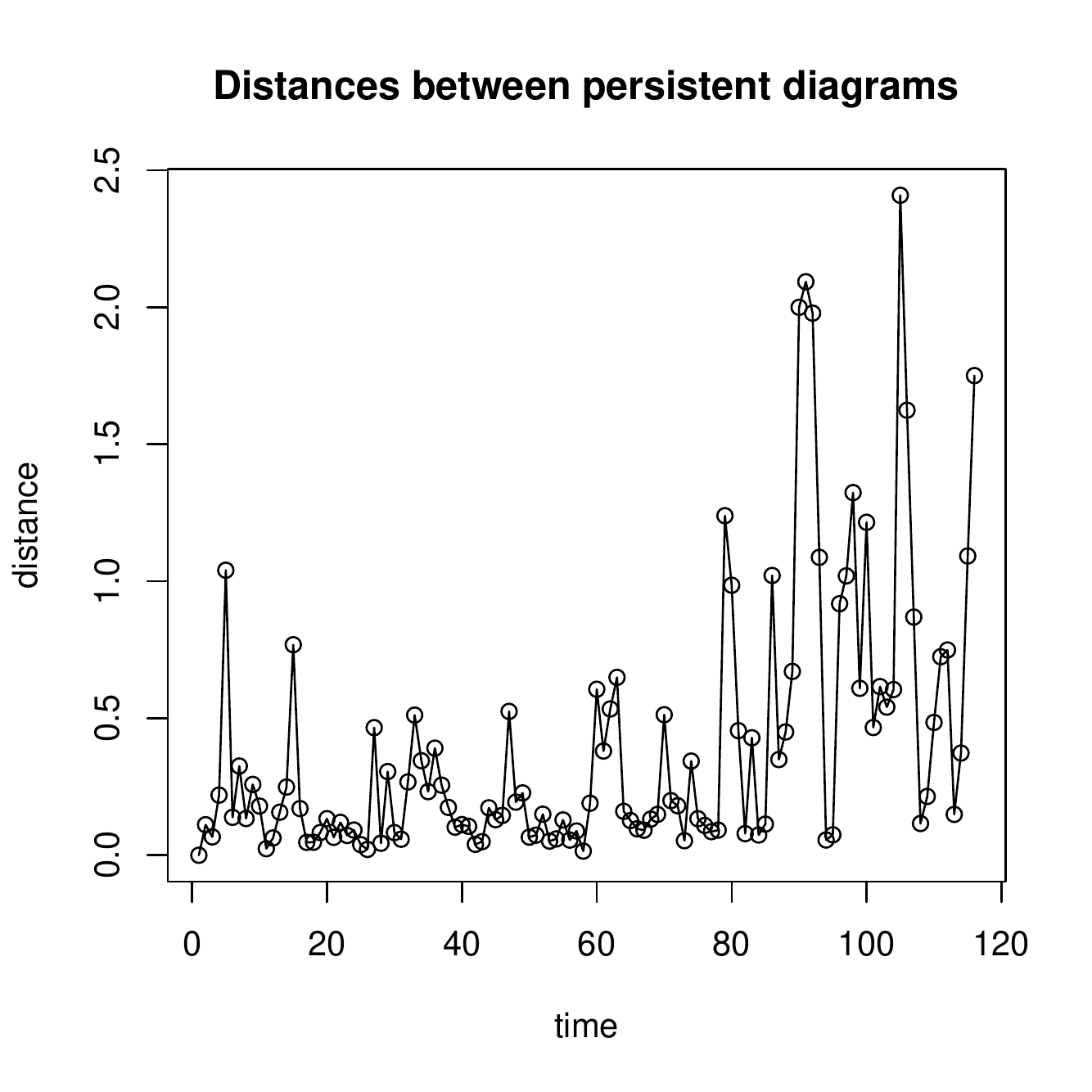}& \includegraphics[width=0.5\textwidth]{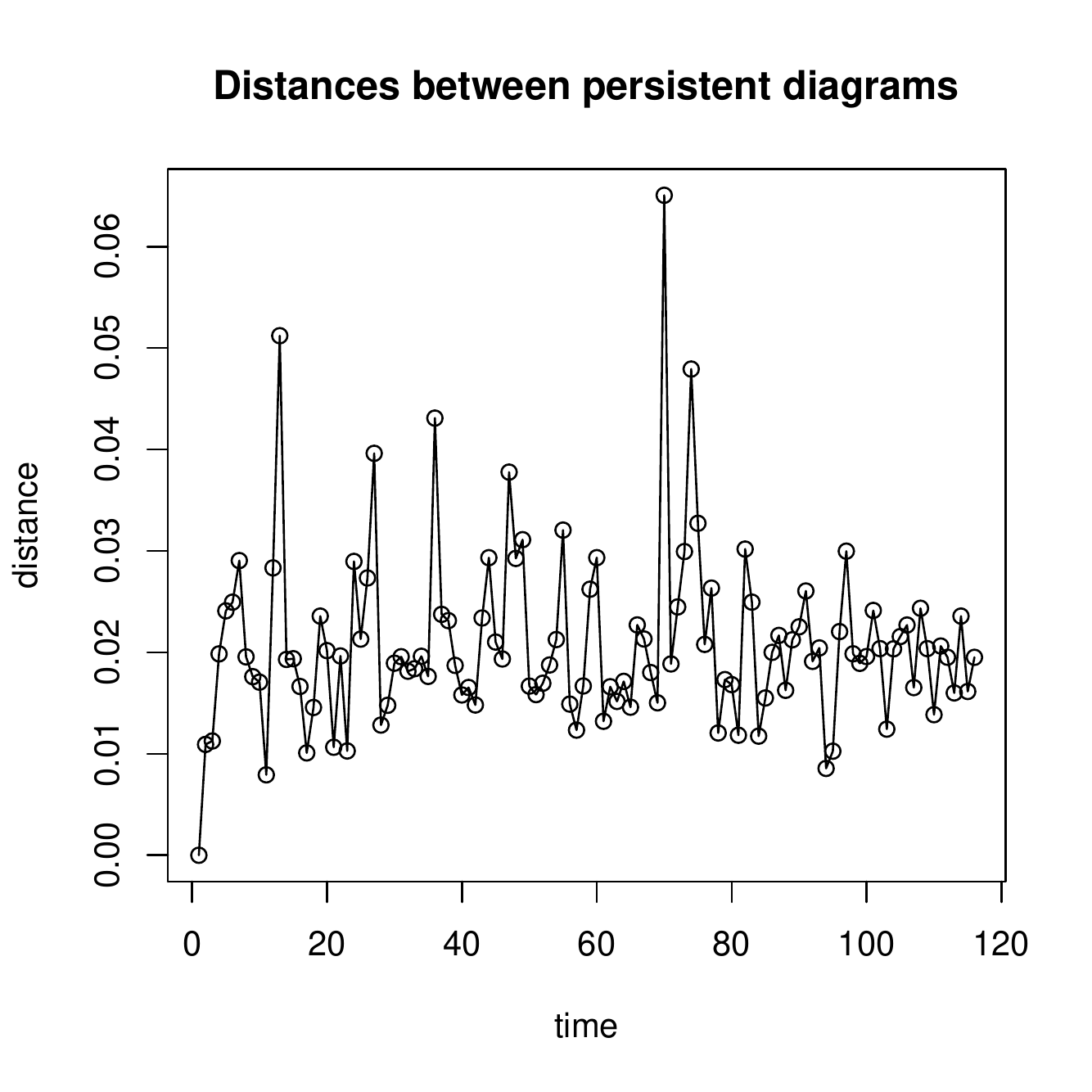}
\end{array}$
\caption{Left: distances between  0-dimensional persistent diagrams  (sub-level sets). Right: Distances between  1-dimensional persistent diagrams  (sub-level sets)}
\label{fig:sublevel-distances}
\end{figure}

These observations can be quantified by computing the time-series representing the distances between the diagram at  time $t$ and some reference persistent diagram at the initial time $t_0$. We sample this time series at $\Delta t=10$. We show this in Fig.~\ref{fig:sublevel-distances}. We use the Wasserstein distance of degree $p=2$. We notice that the oscillations in the second half of the time series are almost double in size when compared with  those in the first half. This shows  a change  in the topology of the network, in terms of its connectivity, when approaching the critical transition.

Now we interpret the $1$-dimensional persistent homology, illustrated  in Fig.~\ref{fig:sublevel-diagrams}
by red marks. The horizontal coordinate of  a mark gives the birth value of a loop in the network, and the vertical coordinate gives the death value of that loop. The death of a loop occurs when edges between the nodes of the loop appear and form complete $2$-simplices (filled triangles) that fill up the loop. Dots with low coordinates indicate the presence of cliques that are highly correlated. Marks with higher vertical coordinates indicate the death of loops due to edges between low-correlated or anti-correlated  nodes. The top three
diagrams in Fig.~\ref{fig:sublevel-diagrams} seem to indicate a concentration of dots at a higher range of values
when compared with the bottom three diagrams.

We also compute the  time-series (sampled at $\Delta t=10$) of the  Wasserstein distances  of degree $p=2$, between the diagram at time $t$ and the reference diagram at $t_0$. We show this in Fig.~\ref{fig:sublevel-distances}. The oscillations in the second part of the time series are smaller  than the ones in the first part. Again, there is a change  in the topology of the network, in terms of its cliques, when approaching the critical transition: the number  of loops of correlated nodes appears to stabilize itself.

\begin{figure}
\centering\[\begin{array}{ccc}
\includegraphics[width=0.3\textwidth]{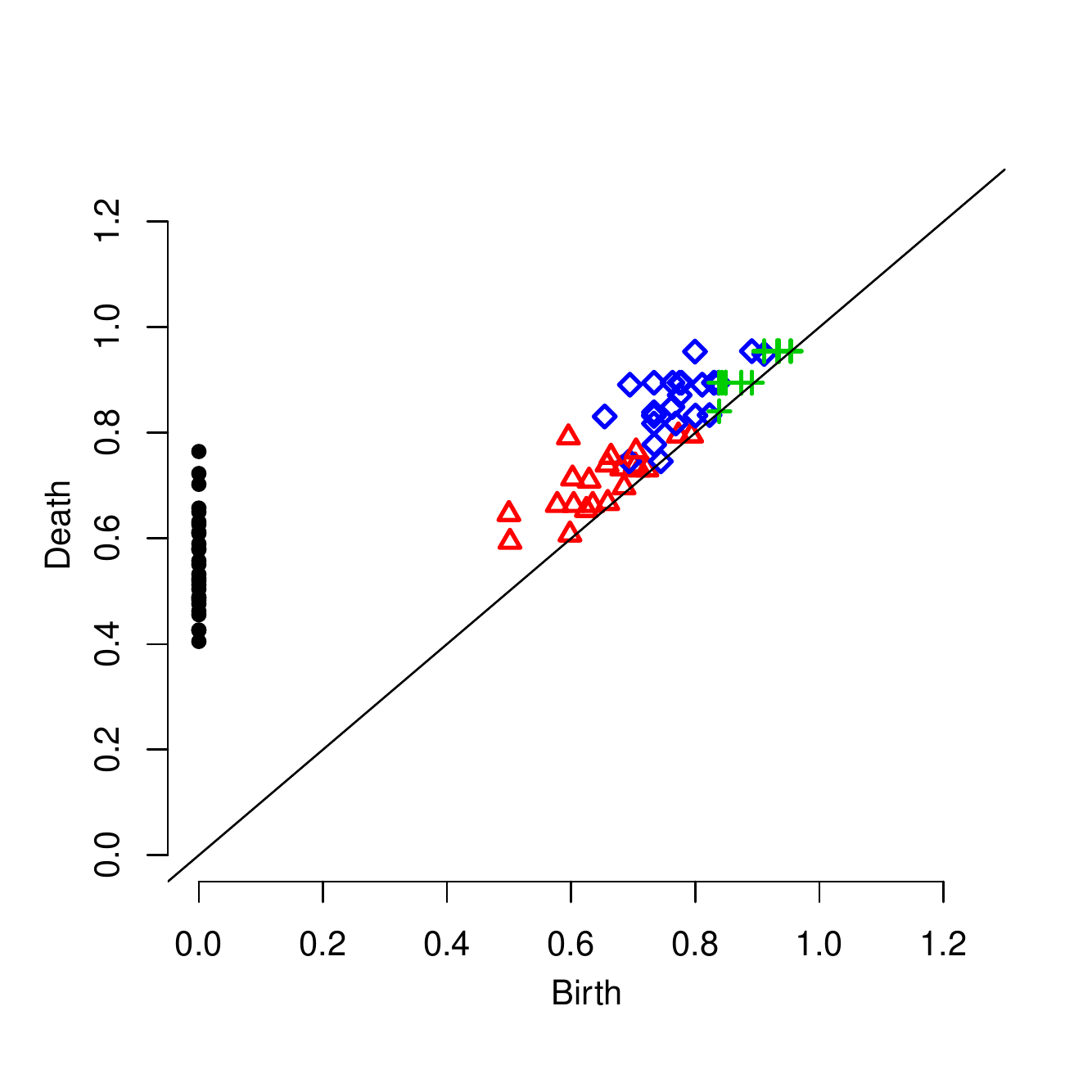}&
\includegraphics[width=0.3\textwidth]{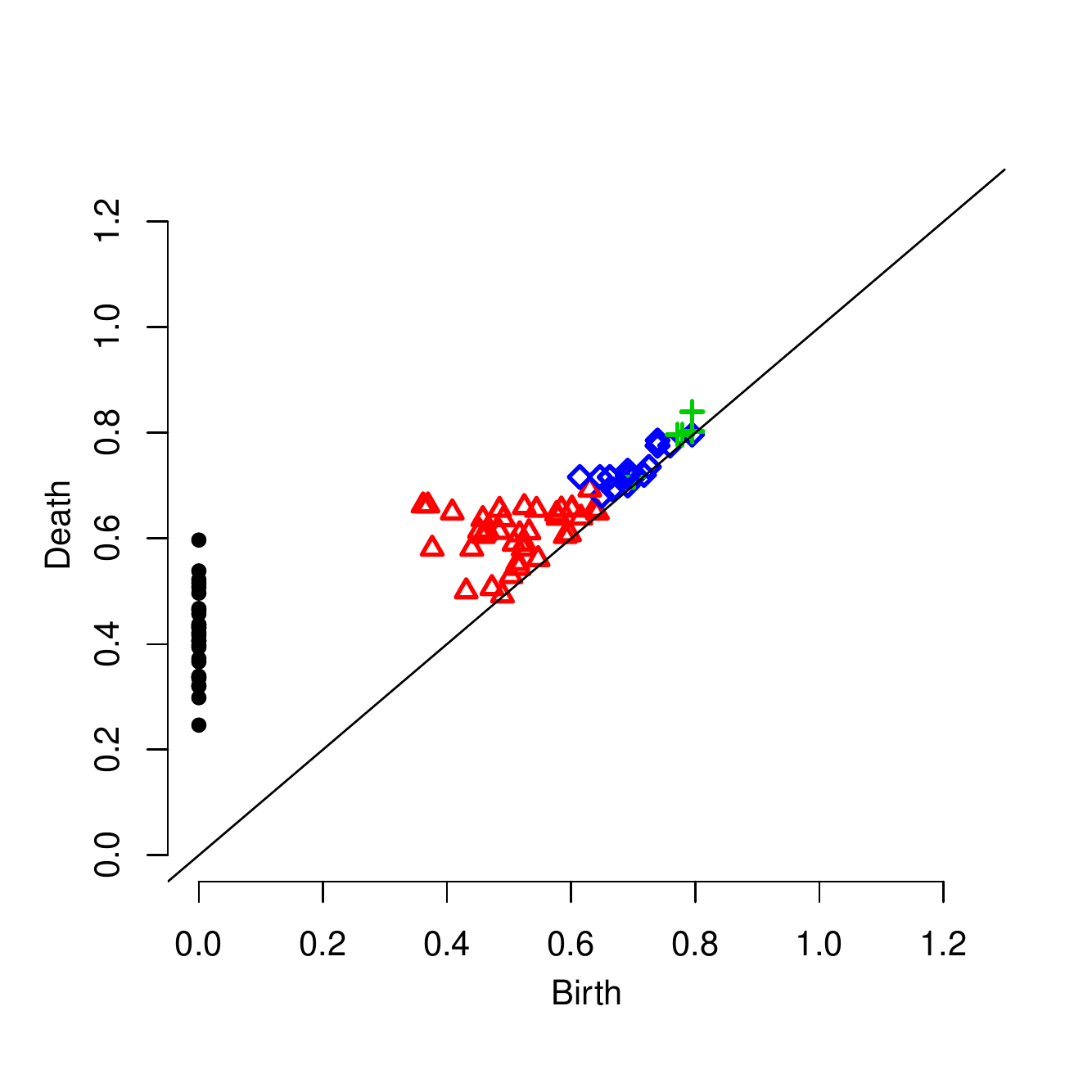}&
\includegraphics[width=0.3\textwidth]{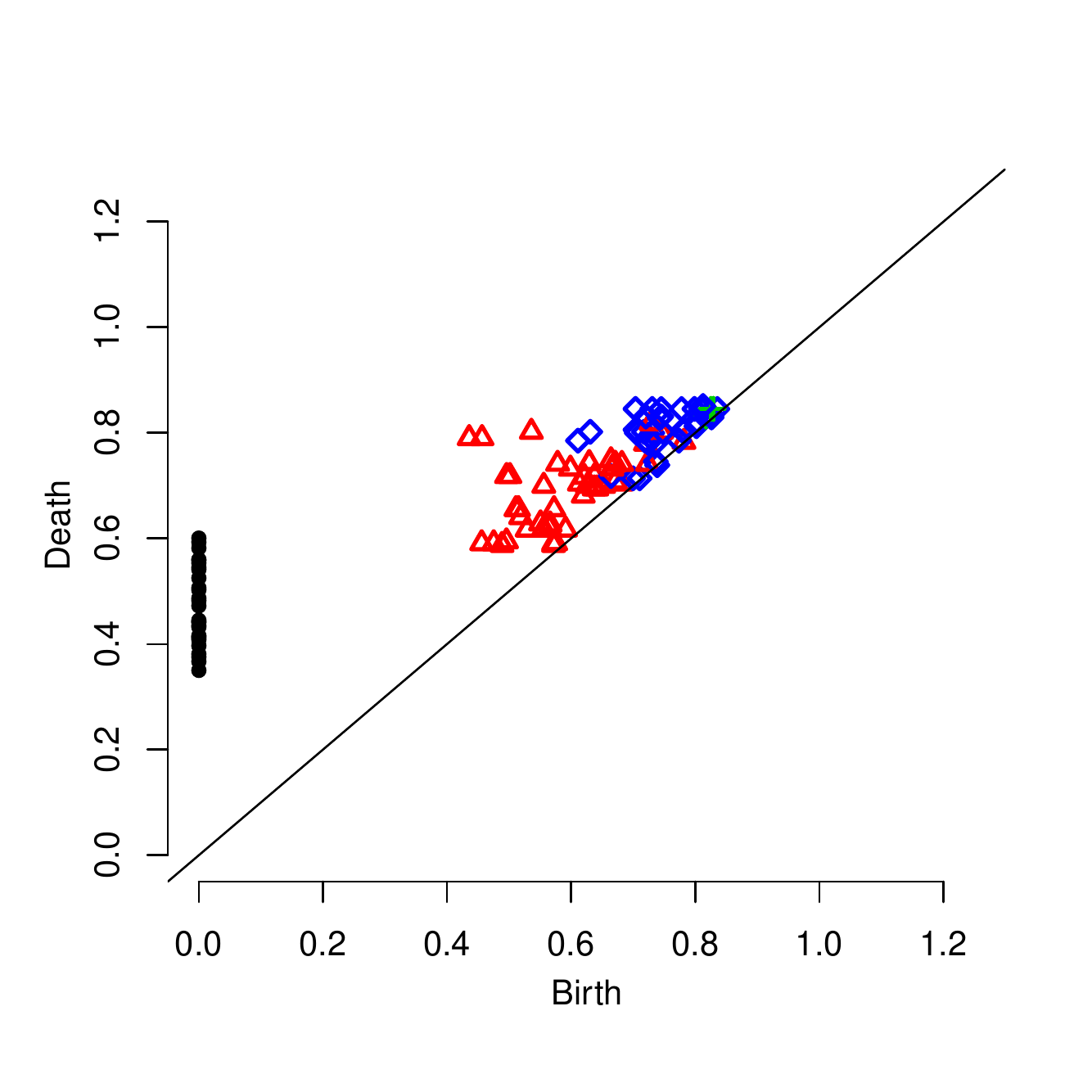}\\
\includegraphics[width=0.3\textwidth]{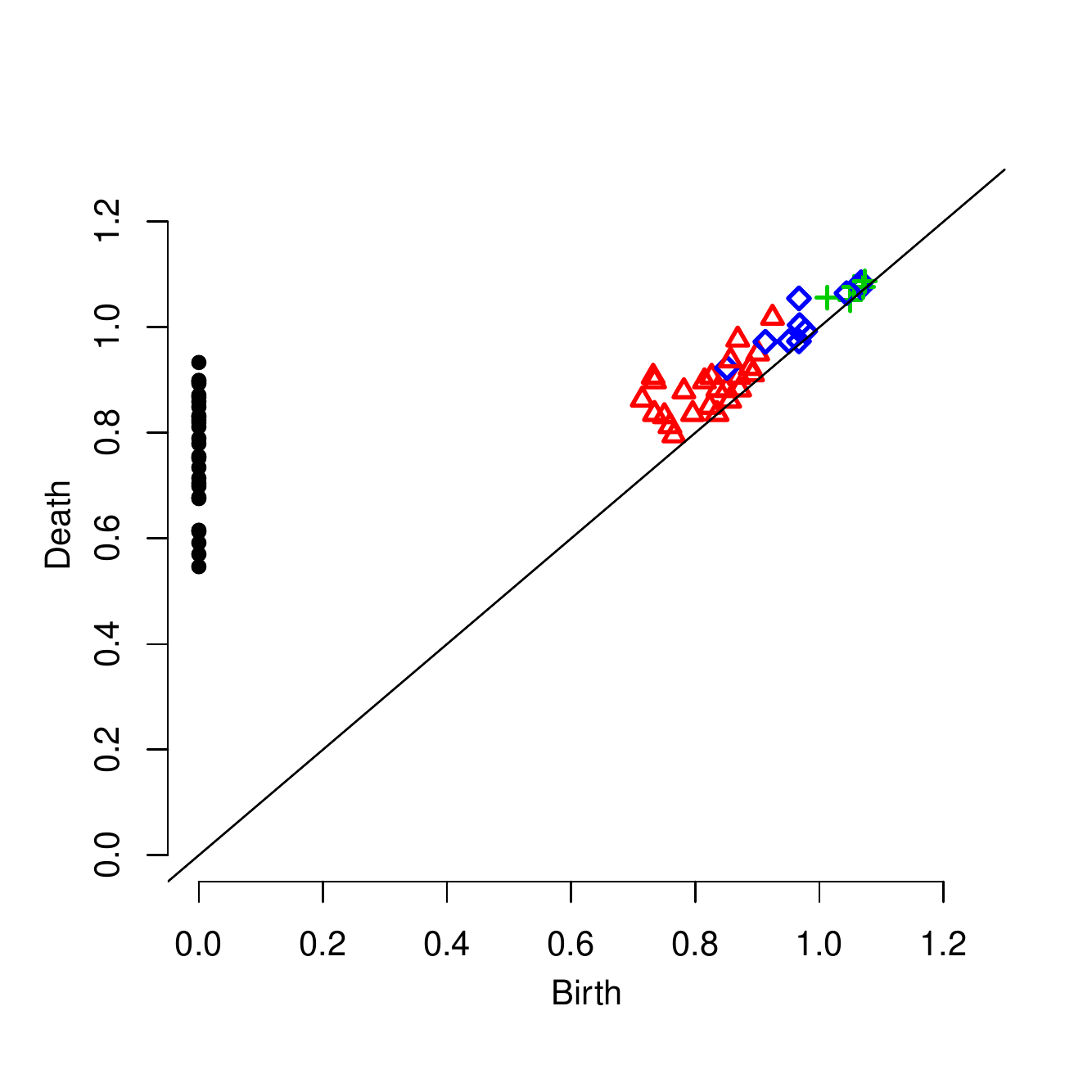}&
\includegraphics[width=0.3\textwidth]{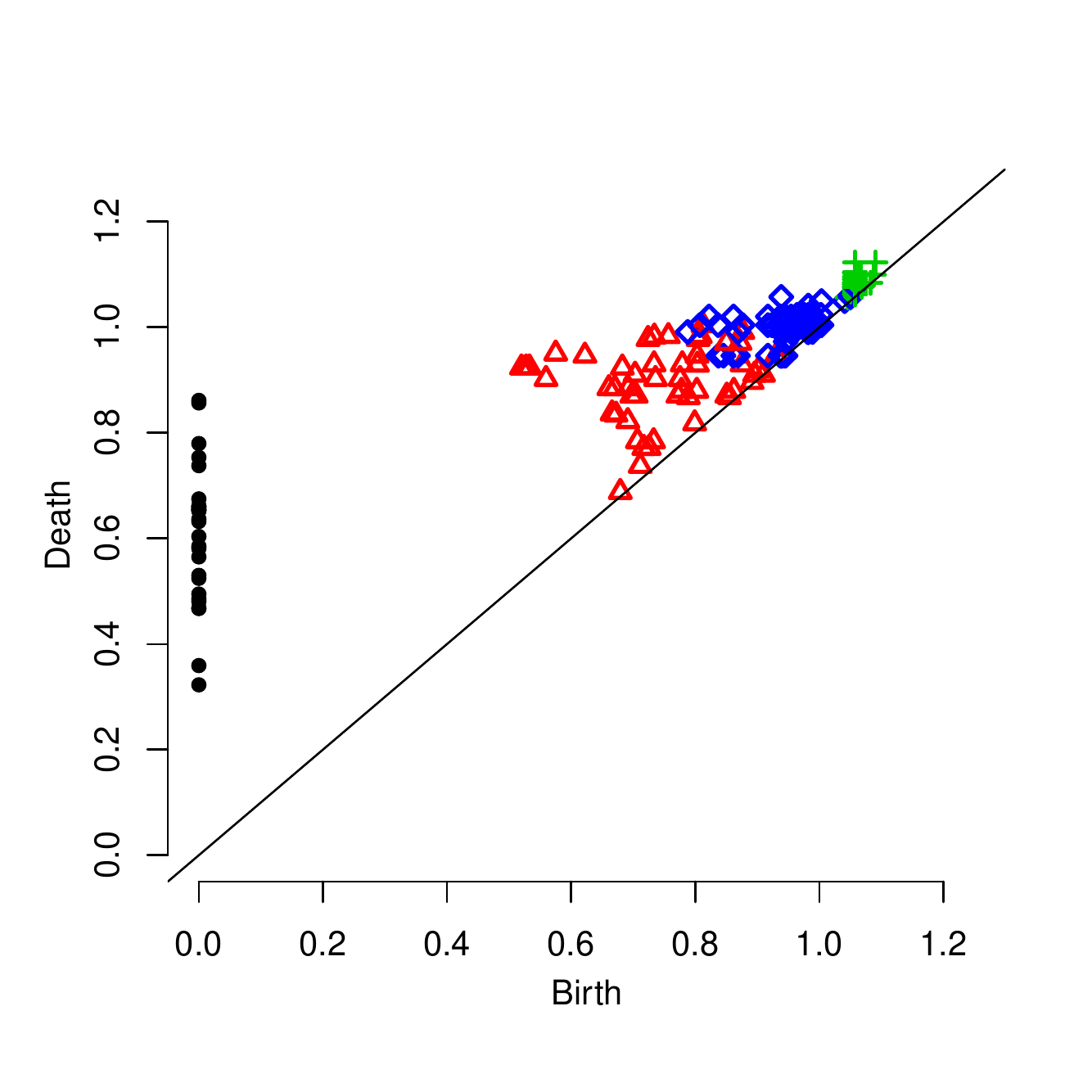}&
\includegraphics[width=0.3\textwidth]{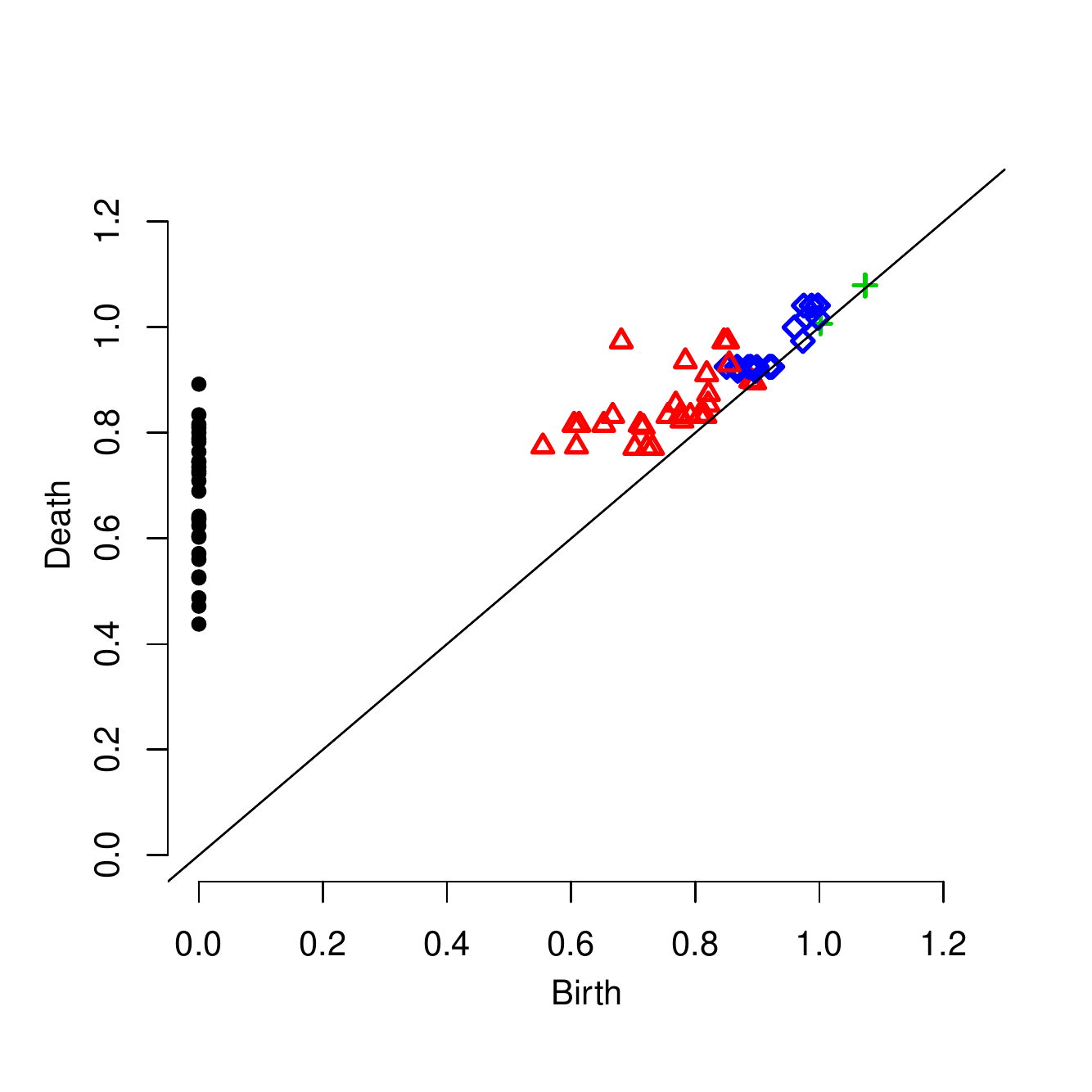}
\end{array}\]
\caption{Persistent diagrams (super-level sets)}
\label{fig:superlevel-diagr}\end{figure}

\begin{figure}
$\begin{array}{cc}
\includegraphics[width=0.5\textwidth]{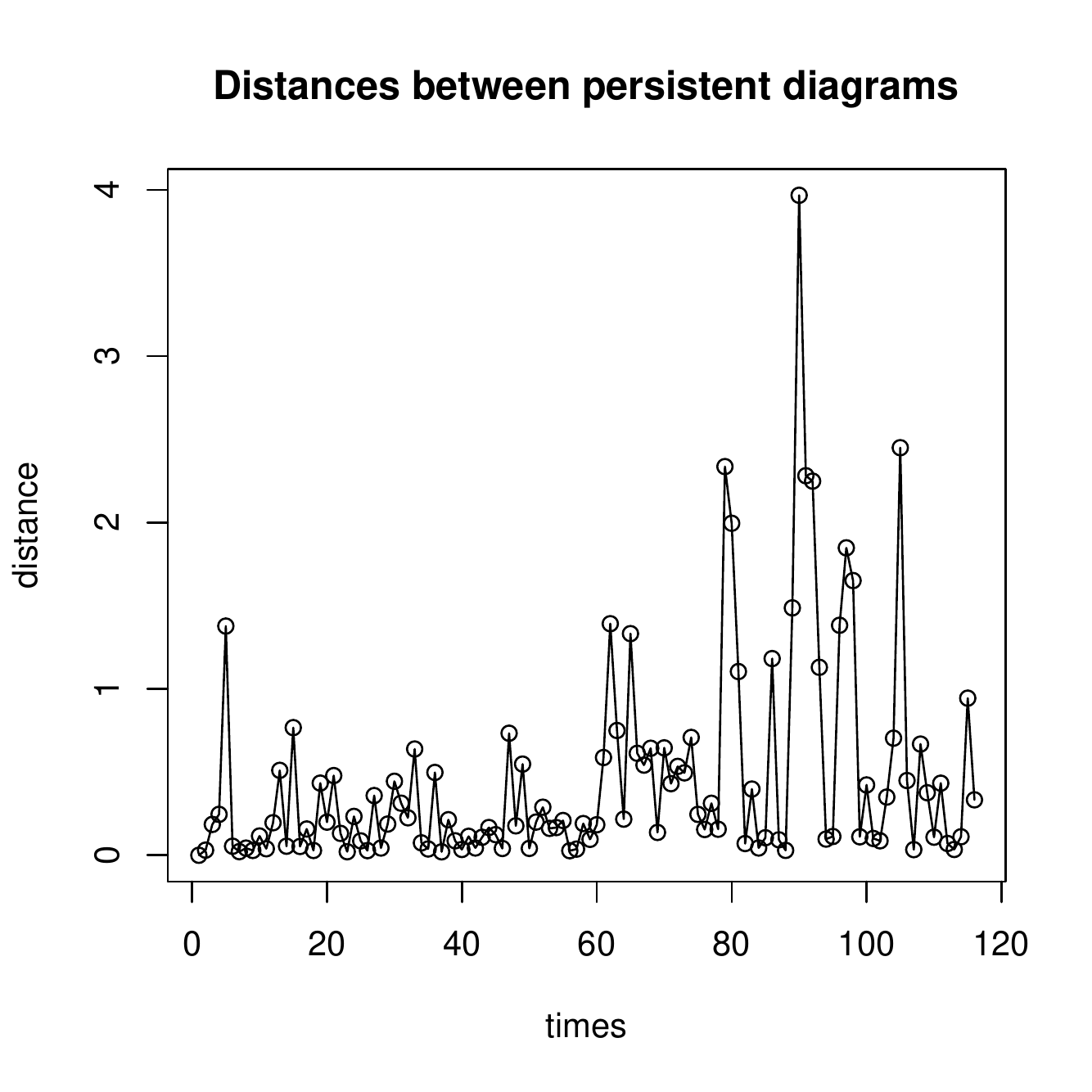}
&
\includegraphics[width=0.5\textwidth]{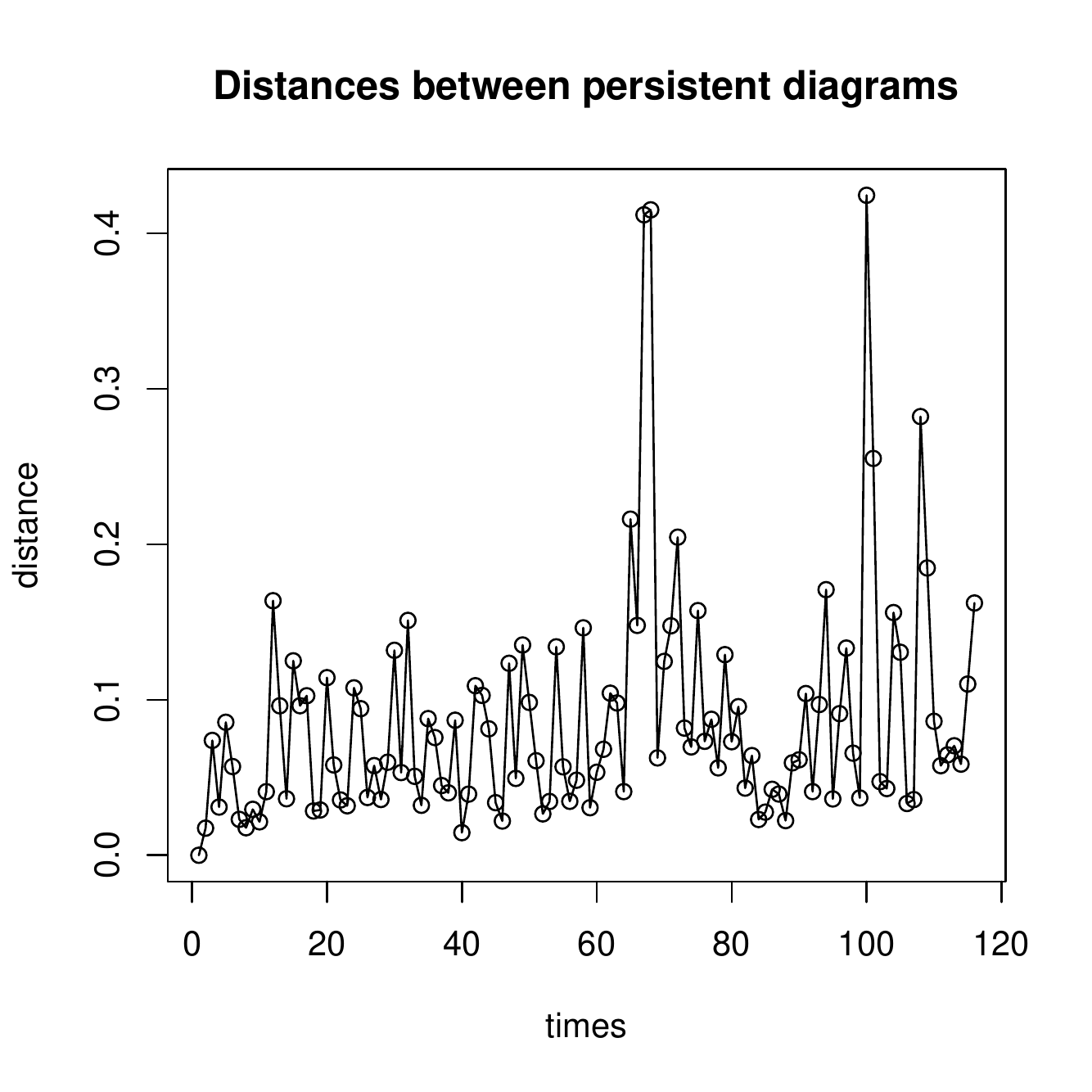}
\end{array}$
\caption{Left: distances between  0-dimensional persistent diagrams (super-level sets). Right: distances between  1-dimensional persistent diagrams  (super-level sets).}
\label{fig:superlevel-distances}
\end{figure}

We now compute the super-levels sets of $w$, which are sub-level set of $w'$. The resulting  persistent diagrams   have a different interpretation. The critical value of the threshold $\theta$ for the switch from anti-correlation to correlations is $0.5857864$. Points in the persistent diagram with low vertical coordinates correspond to anti-correlation/non-correlation, and
points with higher value of the vertical coordinate (other than  $2$) indicates the appearance of edges between correlated nodes. A point on the persistent diagram with higher vertical coordinate represents the  death of a connected component (or a loop), possibly formed by anti-correlated or low correlated nodes,  when an edge between correlated nodes is added to the networks. Thus, the homology generators identified by the persistent diagrams represent cliques of stocks associated to `normal' market conditions (which are associated to lack of correlation). The death of these generators is caused by the addition to correlated edges to the threshold network (in dimension $0$, by joining together different connected components,  and in dimension $1$ by closing the loops). That is, the persistence diagrams capture the loss of normal market conditions. We show some persistent diagrams in Fig.~\ref{fig:superlevel-diagr},  the time series of distances between $0$-dimensional persistent diagrams, and between $1$-dimensional persistent diagrams,  in Fig.~\ref{fig:superlevel-distances}.

\section{Conclusions}
The analysis of the persistent diagrams and of the distances between persistent diagrams show significant changes
in the topology of the correlation network in the period prior to the onset of the 2007-2008 financial crisis;    early signs become apparent starting February 2007 (note that the U.S. stock market peaked in October 2007). These topological changes can be characterized by an increase in the cross correlations between various stocks, as well as by the emergence  of sub-networks of cross correlated stocks.

These results are in agreement with  other research asserting that crises are typically associated with rapid changes in the  correlation structure and in the network topology (see, e.g., \cite{Nobi2014a,Nobi2014b,Munix,Smerlak}).

In addition to the experiments presented here, we have used persistent homology  to analyze the time-series of  some market indices (e.g., the VIX index) for the same period, using point-cloud data sets obtained via delay-coordinate  reconstruction method.   These tests also show early signs of critical transition; those results will be presented elsewhere.

\section*{Acknowledgements}
Research of M.G. was partially supported by  the Alfred P. Sloan Foundation grant G-2016-7320,  and by the NSF grant  DMS-0635607.

\input{NetSciX_Gidea-ref}

\end{document}

%% file: NetSciX_Gidea-ref.tex
%
%